\documentclass[a4paper,11pt]{article}
\usepackage{a4wide}
\usepackage{amsmath}
\usepackage{subcaption}
\usepackage{cite}
\usepackage[normalem]{ulem}
\usepackage{amssymb}
\usepackage{epsf}
\usepackage{amsfonts}
\usepackage{graphicx}
\usepackage{tikz}
\usepackage{pgfplots}
\usepackage{braket}
\usepackage{ragged2e}
\usepackage{appendix}
\numberwithin{equation}{section}
\usepackage{bbm}
\usepackage{xcolor}
\usepackage[export]{adjustbox}
\usepackage{siunitx,pslatex}
\usepackage{booktabs}
\usepackage{hyperref}
\usepackage{ulem}
\usepackage{caption}
\usetikzlibrary{decorations.pathmorphing, decorations.markings, arrows.meta, calc}
\usepgfplotslibrary{fillbetween}
\DeclareMathOperator{\sech}{sech}
\hypersetup{
    colorlinks,
    linkcolor={green!50!black},
    citecolor={magenta},
    urlcolor={blue!80!black}
}
\graphicspath{{Figures/}}

\begin{document}
	
	\title{Comparative study of the butterfly velocity in holographic QCD models at finite temperature and chemical potential}

	\author{\textbf{Nikesh Lilani}\thanks{imnikeshlilani@gmail.com},~ \textbf{Dilpreet Sandhu}\thanks{dilpreetsandhu2255@gmail.com},~ \textbf{Subhash Mahapatra}\thanks{mahapatrasub@nitrkl.ac.in}
		\\\\\textit{{\small Department of Physics and Astronomy, National Institute of Technology Rourkela, Rourkela - 769008, India}}\\
	}
    \date{}
\maketitle

\begin{abstract}
In this work, we study quantum chaos in a variety of holographic QCD models at finite temperature and chemical potentials. This includes the 1 R-Charge black hole (1RCBH) model, the 2 R-Charge black hole (2RCBH) model, a potential reconstruction-based analytic bottom-up model, and a numerical bottom-up model. All these models are different avatars of the Einstein-Maxwell-dilaton gravity action, distinguished by their specific choices of dilaton potentials and gauge-kinetic coupling functions.  We focus on computing the chaos parameter, the butterfly velocity, using three distinct methods: entanglement wedge reconstruction, out-of-time-ordered correlators (OTOCs), and pole-skipping. We show that all three methods yield identical results for the butterfly velocity across all the holographic QCD models considered, further establishing the equivalence between the three approaches. Furthermore, we analyze in detail the behavior of the butterfly velocity as a function of chemical potential and temperature. Interestingly, a universal trend emerges across all models: the butterfly velocity increases/decreases with temperature/chemical potential for thermodynamically stable phases. Additionally, in the high-temperature limit, the butterfly velocity in all models approaches that of the chargeless plasma.
\end{abstract}

\section{Introduction}
\label{sec:intro}
Chaos is an inherent characteristic of a broad spectrum of physical systems, typically demonstrating sensitivity to initial conditions. As a multidisciplinary field within contemporary science, chaos theory employs a variety of analytical tools that have found applications across diverse domains, from the microscopic quantum realm \cite{Srednicki:1994mfb} to the macroscopic structures of spacetime \cite{Suzuki:1996gm}. This theoretical framework enhances our understanding of nonlinear systems that are too intricate for analytical study and allows for the prediction of their long-term behavior. In quantum many-body systems, chaos can be quantified by the growth of out-of-time-ordered correlators (OTOCs)~\cite{Natsuume:2019sfp}.  At early times, a typical OTOC grows as $C(t,\vec x)\simeq 1-\exp[\lambda(t-|\vec x|/v_B)]$, defining a Lyapunov exponent $\lambda$ and a butterfly velocity $v_B$.  A universal bound $\lambda\le2\pi T$ holds at temperature $T$ (the Maldacena–Shenker–Stanford bound)~\cite{Maldacena:2015waa}, and holographic black holes are believed to saturate this bound, acting as maximally chaotic systems.  Such results indicate that black holes can serve as valuable toy models for understanding quantum chaotic dynamics.

While chaos has been extensively studied in various classical and quantum systems, exploring its manifestation in quantum chromodynamics (QCD) poses unique challenges. Unlike simpler systems, QCD generally operates in a regime where the coupling becomes strong, rendering traditional perturbative methods ineffective. This nonperturbative nature complicates the direct calculation of chaotic indicators such as the Lyapunov exponent and butterfly velocity. Moreover, the confinement of quarks and gluons prevents straightforward observation of chaos signatures in experiments. Recent developments, however, have opened new pathways. The gauge/gravity duality, specifically in the context of holographic QCD, has provided an alternative nonperturbative framework to probe chaotic dynamics in the strongly coupled domain of QCD. In this framework, features of quantum chaos, including out-of-time-order correlators (OTOCs) and pole-skipping phenomena, have been successfully mapped to gravitational dynamics in higher dimensions, offering deep insights into the chaotic properties of QCD-like theories~\cite{Hashimoto:2019bl, Giataganas:2018ani}. These developments not only help bridge theoretical gaps but also bring us closer to identifying universal chaotic properties in QCD that could be probed experimentally.

The AdS/CFT correspondence or the gauge-gravity duality~\cite{Maldacena:1997re, Gubser:1998bc, Witten:1998qj} is an equivalence between a gravitational theory on a higher-dimensional Anti-de Sitter (AdS) spacetime and a quantum field theory (QFT) living at its boundary in one lower dimension. This duality, originally envisioned in the context of string theory, implies that a strongly coupled quantum field theory, for example, QCD, can be mapped to a classical gravitational theory.  This correspondence has been remarkably successful in addressing problems in diverse areas of physics, such as quantum gravity, strongly coupled systems, quark-gluon plasma (QGP), etc, to name just a few. It has provided a novel framework for testing phenomena such as confinement, phase transitions, and quantum chaos in gauge theories, thereby providing an alternate technique to probe QCD behavior in the strong coupling domain and thus nonperturbative QCD. One may refer to \cite{Casalderrey-Solana:2011dxg, Rougemont:2023gfz} for detailed reviews on this topic. 

Courtesy of the works \cite{Shenker:2013pqa, Roberts:2014isa, Roberts:2016wdl, Perlmutter:2016pkf}, the study of chaos in strongly coupled many-body quantum systems through the lens of gauge/gravity duality has witnessed significant interest. For a comprehensive review of holographic chaos, one may refer to \cite{Jahnke:2018off}. It was suggested that black holes are the fastest scramblers in nature \cite{Sekino:2008he}. In the context of gauge/gravity duality, a thermal state of the boundary conformal field theory (CFT) corresponds to a black hole in the bulk AdS space. Thus, a rapid thermalization of a local perturbation in the boundary CFT can be understood through the fast scrambling dynamics of the black hole in the dual bulk space. It is also known that black holes in holography can be characterized via quantum chaos \cite{Maldacena:2001kr}.

The gauge/gravity duality has also been used extensively to probe chaotic features of QCD and explored in multiple works. In \cite{Akutagawa:2018yoe, Akutagawa:2019awh}, chaos in the hadronic phase of QCD was studied, with a focus laid on homogeneous meson condensates and time-dependent Wilson loops.  The investigation of chaos in QCD chiral condensates in $\mathcal{N}=2$ supersymmetric QCD was done in ~\cite{Hashimoto:2016wme}, and the $\mathcal{N}=2$ theory was found to be chaotic. In \cite{Hashimoto:2018fkb, Shukla:2023pbp, Shukla:2024qlf, Colangelo:2021kmn, Colangelo:2020tpr}, anisotropic and frame-dependent chaos of a suspended string, corresponding to a quark-antiquark pair in the dual boundary theory, in the holographic deconfined phase was studied. Recently, \cite{Li:2024yma} showed that out-of-time-ordered correlators (OTOC) could also be used to probe the baryonic phase structure in holographic QCD with instantons. Similarly, tools like Krylov complexity and pole-skipping have also been suggested as order parameters for the deconfinement transition \cite{Anegawa:2024wov, Baishya:2023ojl}. By now, a lot of work has been done by analyzing open and closed string dynamics to explore chaos in QCD confining and deconfined phases, for instance, see \cite{Shukla:2024wsu, PandoZayas:2010xpn, Basu:2012ae, Ishii:2016rlk, Basu:2016zkr, Basu:2011di, Banerjee:2018twd,  Xie:2022yef, Ma:2022tvs, Penin:2024rqb, Rigatos:2020hlq, Basu:2011dg}. 

The study of chaotic dynamics in QCD via holography is motivated in part by recent proposals to measure quantum chaos experimentally, notably through OTOCs \cite{Swingle:2016var, Zhu:2016uws, Li:2016xhw, Yao:2016ayk}.  If universal features of chaos in QCD can be captured holographically, they may become accessible to experimental verification. In this work, we extend the analysis of chaotic dynamics in holographic QCD with the aim of identifying such universal signatures across different models.

Among the various hallmarks of chaos, the butterfly velocity ($v_B$) has emerged as a critical quantity that characterizes the spatial spread of perturbations in quantum many-body systems. Conceptually, $v_B$ determines the ``speed limit'' at which information and chaotic disturbances propagate through a system, echoing the butterfly effect in a spatially extended context. In holographic theories, $v_B$ is closely tied to the near black hole horizon dynamics, where it manifests through shockwave geometries and pole-skipping features in retarded correlation functions~\cite{Blake:2016universal, Grozdanov:2019hydro}. Its universality has made it a valuable diagnostic tool in probing the chaotic structure of strongly coupled systems, including QCD-like theories. Notably, $v_B$ has been utilized to explore phase transitions and critical points \cite{Ling:2017butterfly} and the anisotropic nature of quark-gluon plasma via holography~\cite{ Gursoy:2021interplay}. More broadly, $v_B$ has been related to experimentally relevant observables of QGP: recent work showed that both the heavy-quark drag force and the jet-quenching parameter in a thermal plasma are directly controlled by the butterfly velocity~\cite{Ageev:2021poy}. Such findings suggest that chaotic dynamics may manifest in transport and energy-loss phenomena in strongly coupled QCD matter. This makes it not only a theoretical construct but also a potential observable, connecting chaotic dynamics with experimentally measurable transport phenomena.

In this work, our main aim is to investigate the butterfly velocity in the deconfined phase of QCD using holography and study the influence of finite chemical potential $\mu$ and temperature $T$ on QCD chaotic dynamics. These parameters are essential variables in the QCD phase diagram and qualitatively alter its behavior, and we particularly aim to study their effects on the butterfly velocity in various top-down and bottom-up holographic QCD models. The top-down models, while firmly grounded and can be constructed from higher-dimensional string theory and thus are more robust regarding the validity of the duality, often fail to capture key features of QCD. In contrast, bottom-up phenomenological models, typically constructed in an ad hoc manner with limited string-theoretic justification, are more successful in reproducing many desirable aspects of QCD. To make our analysis more concrete and thorough, we consider both top-down and bottom-up holographic QCD models. This includes top-down 1 R-charge black hole (1RCBH) and 2RCBH holographic models \cite{DeWolfe:2011ts, DeWolfe:2012uv}, potential
reconstruction-based analytic bottom-up model \cite{Dudal:2017max, Bohra:2019ebj}, and a numerical bottom-up model \cite{DeWolfe:2010he, Rougemont:2015wca}. All these models are based on the Einstein-Maxwell-dilaton gravity action, distinguished by distinct forms of the dilaton potential and the gauge-kinetic coupling function, unique in their own rights, and have been thoroughly used in the AdS/QCD literature to probe and explore various QCD-related features from holography.

For each of these holographic models, we compute the butterfly velocity using three distinct methods — entanglement wedge reconstruction, OTOCs, and pole skipping — and find that all three yield the same result. Note that such an equivalence between these methods has been previously shown in higher-order curvature gravity theories \cite{Dong:2022ucb}, Lifshitz theories \cite{Baishya:2024gih}, hyperscaling-violating Lifshitz theories \cite{Lilani:2024fth}, $T\bar{T}$ deformed setup \cite{Basu:2025exh}, etc., among various other holographic theories. For other related works, see \cite{DiNunno:2021eyf, Fischler:2018kwt, Eccles:2021zum, Saha:2024bpt,Chakrabortty:2022kvq, Karan:2023hfk}. Our work, therefore, establishes the equivalence between these three methods in AdS/QCD models. Note that the conceptual explanation for the equivalence between the three methods was presented recently in \cite{Chua:2025vig}. Importantly, for the first three models, we obtain analytic results for the butterfly velocity, whereas for the fourth model, we obtain results numerically. We further analyze the thermal- and chemical-dependent profile of the butterfly velocity and find that it exhibits similar features across models. In particular, it increases and then saturates to conformal values as the temperature increases, while it decreases with the chemical potential, suggesting an intricate interplay between temperature, chemical potential, and the butterfly velocity. 

The structure of the paper is as follows. In Section \ref{butterflyvelocity}, we discuss in detail the three methods, namely entanglement wedge, OTOC, and pole skipping, to compute butterfly velocity $v_{B}$. We consider a very general form of planar and isotropic metric and arrive at a general expression for $v_{B}$ using each of the three methods. In Section \ref{subsec:RCBH}, we compute the butterfly velocity in 1RCBH and 2RCBH holographic models. In Sections \ref{subsec:ourmodel} and \ref{subsec:Bottomu2p}, we repeat the butterfly computation in bottom-up holographic QCD models of \cite{Dudal:2017max} and \cite{Rougemont:2015wca}, respectively. Finally, in Section \ref{sec:conclusion}, we conclude and discuss future directions to extend our work.

\section{Operator size and the butterfly velocity}
\label{butterflyvelocity}

\subsection{Entanglement wedge}
\label{sec:EW}
Using the tools developed in~\cite{Mezei:2016wfz}, we can compute the butterfly velocity in holographic QCD systems from the bulk entanglement wedge. An entanglement wedge is essentially a subregion in the bulk that is bounded by a certain subregion on the boundary and whose extremal surface is homologous to the boundary subregion. It is a codimension-zero bulk region that is naturally associated with the given boundary spatial region and, hence, with the associated reduced density matrix. More details on the entanglement wedge and the subregion-subregion duality can be found in \cite{Czech:2012bh, Wall:2012uf, Headrick:2014cta, Dong:2016eik}.

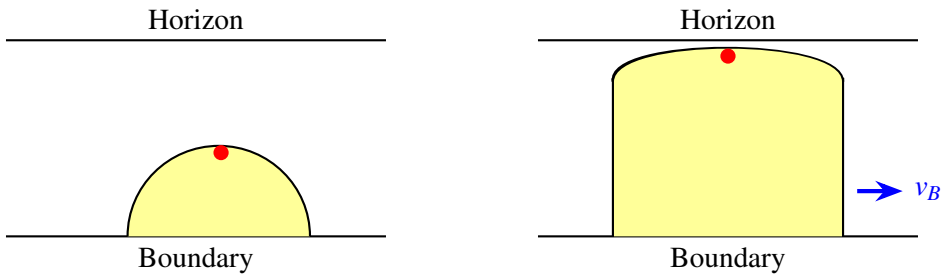
\begin{figure}[h!]
\centering
\begin{tikzpicture}
\definecolor{yellowfill}{RGB}{255, 255, 153}
\definecolor{reddot}{RGB}{255, 0, 0}
\draw[thick] (-5,2.6) -- (0,2.6) node[midway, above] {Horizon};
\draw[thick] (-5,0) -- (0,0) node[midway, below] {Boundary};
\draw[thick] (2,2.6) -- (7,2.6) node[midway, above] {Horizon};
\draw[thick] (2,0) -- (7,0) node[midway, below] {Boundary};
\fill[yellowfill] (-1,0) arc[start angle=0, end angle=180, radius=1.2] -- cycle;
\draw[thick] (-1,0) arc[start angle=0, end angle=180, radius=1.2];
\fill[reddot] (-2.17,1.11) circle (0.1);
\coordinate (A) at (3,0);
\coordinate (B) at (3,2.05);
\coordinate (C) at (6,2.05);
\coordinate (D) at (6,0);
\draw[ultra thick] (A) to[out=90, in=90, looseness=0.5] (B) to[out=100, in=90, looseness=0.5] (C) to[out=90, in=90, looseness=0.5] (D);
    \fill[yellowfill] (A) to[out=90, in=90, looseness=0.5] (B) to[out=90, in=90, looseness=0.5] (C) to[out=90, in=90, looseness=0.5] (D);
\fill[reddot] (4.5,2.39) circle (0.1);
\draw [ultra thick,blue, arrows = {-Stealth}] (6.2,0.6) -- (6.8,0.6) node[right] {$v_B$};
    \end{tikzpicture}
    \caption{The yellow shaded region represents the entanglement wedge and the red dot represents the particle that originates from the boundary. On the left, entanglement wedge at a certain time $t$ is shown. On the right, entanglement wedge at time $t'$ is shown. $t'$ is an adequately later time when the Ryu-Takyanagi surface reaches very close to the horizon.}
    \label{fig:EW}
\end{figure}

In \cite{Mezei:2016wfz}, the delocalization dynamics of a local operator in a thermal equilibrium state were analyzed through the lens of holographic duality. When a local operator acts on the boundary thermal state, it introduces a perturbation represented in the bulk as a particle propagating toward the horizon of a static black hole. Over time, the spatial support of the operator in the boundary theory expands, requiring increasingly larger boundary subregions to reconstruct information about the perturbation. This growth is governed by the butterfly velocity \(v_B\), which characterizes the effective light-cone speed of operator spreading in chaotic systems. In the bulk description, the smallest boundary subregion capable of reconstructing the perturbation corresponds to an entanglement wedge -- the bulk region bounded by the Hubeny-Rangamani-Takayanagi (HRT) surface anchored to the subregion. As the particle falls inward, the radial depth of its trajectory determines the minimal boundary subregion whose entanglement wedge contains the particle. This establishes a dynamical relationship: the time-dependent spatial extent of the operator on the boundary is dual to the radial infall of the particle in the bulk, with \(v_B\) dictating the rate at which the required boundary subregion expands. Figure~\ref{fig:EW} illustrates this process, where the entanglement wedge (yellow region) associated with a boundary subregion grows to encompass the infalling perturbation (red dot) at successive times.  

This mechanism provides a gravitational explanation for entanglement wedge subregion duality in AdS/CFT, demonstrating how the emergent light-cone structure of chaotic systems, encoded in \(v_B\), directly governs the holographic reconstruction of the bulk locality from boundary data. The saturation of \(v_B\)-dependent bounds on operator growth further underscores the connection between entanglement dynamics, scrambling, and the geometric emergence of causal structure in holographic theories \cite{Roberts:2016wdl, Kim:2016wby}. To compute the entanglement wedge, we start with the planar black hole with the following general form of the metric:
\begin{equation}
    \label{eqn:generalmetric}
    ds^{2} = - h(z)g(z)dt^{2} + \frac{f(z) dz^{2}}{g(z)} + k(z)dx^{i}dx^{i}\,,
\end{equation}
here $g(z)$ is the blackening function and $z$ is the usual radial coordinate.\footnote{Throughout this paper, we will denote the holographic radial coordinate by $z$ instead of $r$. Accordingly, the $z$ coordinate runs from $z=z_h$ (horizon) to $z=\infty$ (asymptotic boundary). We reserve the coordinate $r$ to denote the radial coordinate in the $x^{i}$ plane.} The horizon is defined by the condition $g(z_{h}) = 0$, with $z_{h}$ being the horizon radius. Note that the spatial part of the metric is isotropic and planar. Since the position of the particle is time dependent and not the background geometry itself, the Ryu-Takayanagi (RT) surface~\cite{Ryu:2006bv} is taken as the extremal surface in this case and not the HRT surface~\cite{Hubeny:2007xt}. Thus, for our purposes, the entanglement wedge is a $t = \text{constant}$ hypersurface bounded by the RT surface and the boundary subregion. 

The entanglement entropy or the area functional is given as follows:
\begin{equation}
    \label{eqn:SEE}
    S_{EE} = 2\pi \int \sqrt{\gamma}
    \, d^{d-1}\xi \,,
\end{equation}
where $\xi$ are the appropriate coordinates on the codimension-two surface in the bulk, and $\gamma$ is the determinant of the induced metric on the surface. Extremizing $S_{EE}$ (or, equivalently, the induced metric $\gamma$) gives rise to the equation of the RT surface.  Since the surface is embedded in a $t=\text{constant}$ hypersurface, the $tt$ component of the induced metric vanishes. Further, we take $r$ as the radial coordinate in the $x^{i}$ direction ($ r = |x|$) and parametrize $z$ with $r$, i.e., $z = z(r)$. Hence, from \eqref{eqn:generalmetric} we get the following induced metric:
\begin{equation}
    \label{eqn:inducedmetric}
    \gamma_{ab}d\xi^{a}d\xi^{b} = \left[\frac{(z')^{2}f(z)}{g(z)}+ k(z)\right]dr^{2} + k(z)r^{2}d\Omega_{d-2}^{2}\,.
\end{equation}
Next, analogous to the computation of \cite{Mezei:2016wfz}, we do a near-horizon analysis of the RT surface to compute the butterfly velocity. Near the horizon, we can take the following form of $z(r)$:
\begin{equation}
    \label{eqn:zr} 
    \frac{1}{z(r)} = \frac{1}{z_h} - \epsilon u(r)^{2}\,,
\end{equation}
where $u(r)$ is the function to be determined by extremizing \eqref{eqn:SEE}. Substituting \eqref{eqn:zr} into the induced metric, expanding $\gamma$ near the horizon up to order $\mathcal{O}(\epsilon)$, and using the fact that the blackening function $g(z)$ vanishes at the horizon, we get
\begin{equation}
    \label{eqn:detinduced}
    \sqrt{\gamma} = \sqrt{k_{o}^{d-1}}r^{d-2}\left(1+\epsilon z_h^2 \frac{(4f_{o}(u')^{2} + (d-1)u^{2}g'_{o}k'_{o})}{2k_{o}g'_{o}}\right)\,,
\end{equation}
where the subscript $o$ indicates that the functions are evaluated at $z=z_{h}$ and the prime $'$ indicates the derivative with respect to $z$. Extremizing \eqref{eqn:detinduced}, we arrive at the following differential equation for $u(r)$:
\begin{equation}
\label{eqn:ODEforu}
    u''(r) + (d-2) \frac{u'(r)}{r} - \frac{(d-1)g'_{o}k'_{o}}{4f_{o}}u(r) = 0\,.
\end{equation}
The RT surface starts to depart the near-horizon region for large $r$. To put it precisely, the RT surface stays close to the horizon up to the point where $\epsilon u(r)^{2} \sim$ $\mathcal{O} (\frac{1}{z_h})$, and after that, the surface departs the near-horizon region and reaches the boundary to order one distance. As suggested  in \cite{Dong:2022ucb}, we can encapsulate this behavior via the following ansatz:
\begin{equation}
    \label{eqn:ansatz}
    u(r) \sim \frac{e^{\sigma r}}{r^{n}}\,,
\end{equation}
where $n$ is some positive integer. Also, the particle touches the tip of the RT surface at all times. Thus, taking the tip of the RT surface to be the origin, we can set $u(o,t) \sim e^{-2 \pi t/\beta}$. Thus $u(r,t)$ is given as follows:
\begin{equation}
    \label{eqn:urt}
    u(r,t) \sim \frac{e^{\sigma r - \frac{2 \pi t}{\beta}}}{r^{n}}\,.
\end{equation}
The rate at which the particle propagates toward the horizon is the rate at which the size of the applied operator grows (we make the notion of ``size'' more precisely in Section \ref{sec:OTOC}). Thus, from \eqref{eqn:urt}, the butterfly velocity is given as
\begin{equation}
    \label{eqn:vb}
    v_{B} = \frac{2 \pi }{\beta \sigma}\,.
\end{equation}
To determine $v_{B}$, we need to determine $\sigma$. To determine $\sigma$, we substitute the ansatz \eqref{eqn:ansatz} into \eqref{eqn:ODEforu}. Dropping higher-order terms in $1/r$, we get the following expression for $\sigma$:
\begin{equation}
    \label{eqn:muexpression}
    \sigma^{2} = \frac{(d-1)g'_{o}k'_{o}}{4f_{o}}\,.
\end{equation}
Substituting \eqref{eqn:muexpression} into \eqref{eqn:vb}, we get
\begin{equation}
    \label{eqn:vbT}
    v_{B}^{2} = 16 \pi^{2} T^{2} \frac{f_{o}}{(d-1)g'_{o}k'_{o}}\,.
\end{equation}
It is worth noting that, if the temperature of the black hole is not a function of the chemical potential (like in certain holographic models), the temperature does not become zero for any value of the chemical potential. Also, the function $f(z)$ cannot vanish at the horizon because $g_{o}=0$ [from \eqref{eqn:generalmetric}]. This implies that $v_{B}$ in such models is never exactly zero for any finite value of the chemical potential. We will consider such an example in Section \ref{subsec:RCBH}.

Now, we can also express the temperature of the black hole in terms of the metric functions
\begin{equation}
    \label{eqn:tempbh}
    T = \frac{\sqrt{-(g^{zz})'_{o}(g_{tt})'_{o}}}{4 \pi}  = \sqrt{\frac{h_{o}}{f_{o}}}\frac{g'_{o}}{4 \pi}\,. 
\end{equation}
Substituting \eqref{eqn:tempbh} into \eqref{eqn:vbT}, we get the butterfly velocity completely in terms of the metric functions
\begin{equation}
    \label{eqn:vbfinal}
    v_{B}^{2} = \frac{g'_{o}h_{o}}{(d-1)k'_{o}}\,.
\end{equation}

\subsection{Shock wave and OTOCs}
\label{sec:OTOC}
We are interested in the spreading of the quantum butterfly effect in spatial directions of the boundary QCD systems at strong couplings. The same is intrinsically related to the size of perturbative local operators and can be obtained from the OTOCs of local operators at the boundary. In the dual gravity picture, one can think of this as injecting a small number of quanta, which propagate toward the black hole in the bulk and create a localized shock wave. The spatial region where the shock wave has nontrivial support determines the size of the corresponding boundary operator. Accordingly, one can evaluate the butterfly velocity by computing how this special region grows by sending perturbations at progressively earlier times.

The operator size or the spreading of the quantum butterfly effect in a system can be quantified via the following commutator:
\begin{equation}
 \label{eqn:commutator}
    C(x,t) = \left< - [ W(0,-t),V(x,0)]^{2}\right>_{\beta}\,,
\end{equation}
where $V$ and $W$ are two Hermitian and unitary operators, $\braket{\mathcal{O}}\equiv \text{tr}\left[e^{-\beta H}\mathcal{O}\right]/\text{tr}e^{-\beta H}$, and $\beta=1/T$ is the inverse of the temperature. $C(x,t)$ accounts for the effect that the perturbation $W$ at earlier time $-t$ and position $x=0$ has, on the operator $V$ at time $t=0$ and position $x$. 
We make use of the negative time because it becomes convenient once we switch to the holographic picture. 
This commutator of two spatially separated operators generally behaves like \cite{Shenker:2013pqa, Roberts:2014isa, Roberts:2016wdl} 
\begin{equation}
    \label{eqn:expcommutator}
    C(x,t) \sim \exp{\left[\lambda_{L}\left(t - t_{*}-\frac{|x|}{v_{B}}\right)\right]} \,,
\end{equation}
where $\lambda_{L}$ is the quantum Lyapunov exponent, $t_{*}$ is the scrambling time, and $v_{B}$ is the butterfly velocity. Note that \eqref{eqn:expcommutator} results from a purely boundary calculation. The scrambling time $t_{*}$ is defined as the time at which the commutator with $x=0$ becomes order unity. In chaotic systems, the commutator with $x=0$ exhibits an exponential growth up to the scrambling time. After the scrambling time, the information about a local perturbation starts to scramble among the local degrees of freedom at a constant rate, characterized by the butterfly velocity $v_{B}$. After time $t$ of inserting the perturbation operator $W (|t| > |t_{*}|)$, the commutator is of order unity in the region given by $ |x| < v_{B}|t-t_{*}|$, and this region is said to be the ``size'' of the operator $W$, i.e., it is the region in which any operator is significantly affected by $W$.

The above commutator can be expanded as
\begin{equation}
  \label{eqn:otoc}
    C(x,t) = 2 - 2\left < W(0,-t) V(x,0)W(0,-t) V(x,0)\right> \,.
\end{equation}
The second term in \eqref{eqn:otoc} is what is called the out-of-time-ordered correlator (OTOC), and it entails all the nontrivial information of the commutator. The OTOC is used to diagnose chaos in a quantum system, as the vanishing of the OTOC implies chaotic time evolution of the system. This can be physically understood by considering the following two states:
\begin{eqnarray}
\label{eqn:twostates}
    \ket{\psi_{1}} & = & W(0,-t)V(x,0)\ket{\beta}\,,  \nonumber \\
    \ket{\psi_{2}} & = & V(x,0)W(0,-t)\ket{\beta}\,,
\end{eqnarray}
where, in the context of shock wave analysis, the state $\ket{\beta}$ would correspond to a thermofield double state. The physical interpretation of the state $\ket{\psi_{1}}$ is as follows: At time $t = 0$ (and spatial point $x$), state $V(x,0)\ket{\beta}$ is prepared and then evolved backward in time via the unitary time evolution ($e^{iHt}$, where $H$ is the Hamiltonian). A small perturbation $W$ is made and the system is evolved forward in time again to $t=0$. If we act with $W$ at a sufficiently early time in the past and if the system is chaotic, it would interfere with the perturbation due to $V$, thereby preventing the operator $V$ from rematerializing at $t = 0$. Similarly, the state $\ket{\psi_{2}}$ can be understood as follows: A perturbation $W$ is made at time $-t$ (and at $x=0$), then the system evolves to time $t=0$ at which the operator $V$ is inserted at a spatial point $x$. Since the operator $W$ scrambles among the degrees of freedom in its local region in chaotic systems, the overlap between the two states $\ket{\psi_{1}}$ and $\ket{\psi_{2}}$ becomes small. As a result, the OTOC vanishes in chaotic systems, causing the commutator in Eq.~\eqref{eqn:otoc} to grow.

\subsubsection{Kruskal extension and the shock wave analysis}
In the previous subsection, we discussed OTOC from the boundary perspective. In this subsection, we explore the bulk perspective and discuss how the vanishing of OTOCs can be understood within the bulk. Additionally, we discuss how the butterfly velocity can be computed holographically through shock wave analysis in the bulk. Most of our discussion here is based on \cite{Dong:2022ucb}; more details can be found there.

To understand the bulk picture, we start with the two-sided black hole geometry in the bulk. This corresponds to a thermofield double state $\ket{\text{TFD}}$ at the boundary and is given by \cite{Maldacena:2001kr}
\begin{equation}
\label{eqn:TFD}
|\text{TFD}\rangle_\beta \equiv \frac{1}{Z^{1/2}} \sum_n e^{-\beta E_n / 2} |n\rangle_L |n\rangle_R\,,
\end{equation}
where $Z$ is the thermal partition function, $\{\ket{n}\}$ is a complete set of energy eigenstates, and $E_{n}$ are the corresponding eigenvalues. $L$ and $R$ label the two copies of the states. To work with the two-sided geometry, it is convenient to write the planar black hole metric in Kruskal coordinates. We consider the following Kruskal coordinate transformation:
\begin{equation}
\label{eqn:kruskaltransformation}
    u = e^{\left[\frac{2\pi}{\beta}(z_{*}-t)\right]},
  \hspace{0.4cm}  
  v = -e^{\left[\frac{2\pi}{\beta}(z_{*}+t)\right]}\,,
\end{equation}
where, as usual, $\beta$ is the inverse Hawking temperature and $z_{*}$ is the tortoise coordinate defined as 
\begin{equation}
\label{eqn:tortoise}
    dz_{*} = \frac{dz}{g(z)}\sqrt{\frac{f(z)}{h(z)}}\,.
\end{equation}
In the Kruskal coordinates, the metric takes the form 
\begin{equation}
    \label{eqn:kruskalmetric}
    ds^{2} = 2A(u,v)dudv + B(u,v)dx^{i}dx^{i}\,,
\end{equation}
with $A(u,v)$ and $B(u,v)$ given by
\begin{equation}
    \label{eqn:relationofAandB}
    A(u,v) = \frac{g(z(u,v))h(z(u,v))}{2 \alpha^{2} uv}, \hspace{0.3cm} B(u,v) = k(z(u,v))\,,
\end{equation}
where $\alpha = 2 \pi/\beta$. In these coordinates, the two horizons correspond to $u=0$ and $v=0$. It is important to note that, generally, \eqref{eqn:tortoise} is not integrable. Since most of the computation in shock wave analysis requires near-horizon physics, we can series expand the integral around the horizon. The dominant term in the expansion is given by
\begin{equation}
    \label{eqn:logterm}
    z_{*} \approx \sqrt{\frac{f_{o}}{h_{o}}}\frac{1}{g'_{o}}\ln(z-z_h)\,.
\end{equation}
Also,
\begin{equation}
    \label{eqn:uv}
    uv = - e^{2\alpha z_{*}} = -e^{4 \pi T z_{*}}\,.
\end{equation}
Substituting \eqref{eqn:logterm} and \eqref{eqn:tempbh} into \eqref{eqn:uv}, we get the following:
\begin{equation}
    \label{eqn:zhminusuv}
    z = z_h - uv\,.
\end{equation}
Putting $u=0$ or $v=0$ gives $z=z_h$, which implies that $u=0$ and $v=0$ indeed represent the horizon. 

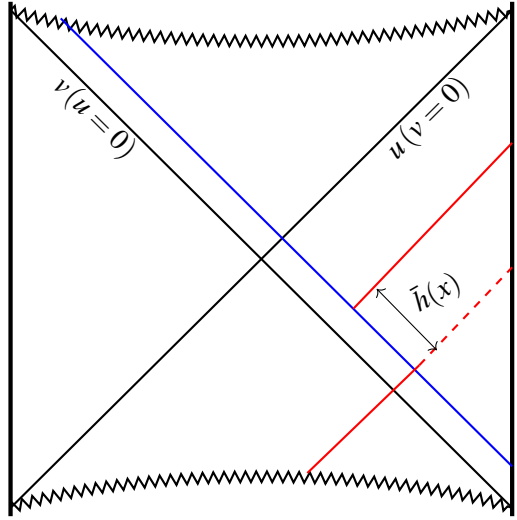
\begin{figure}[h!]
    \centering
    \begin{tikzpicture}[scale=1.1] 
    \draw[ultra thick] (-3.0,-3.1) -- (-3.0,3.1); 
    \draw[ultra thick] (3.0,-3.1) -- (3.0,3.1);  
    \draw[thick,decorate,decoration={zigzag,amplitude=2pt,segment length=4pt}] 
        (-3,3) .. controls (-1.5,2.5) and (1.5,2.5) .. (3,3); 
    \draw[thick,decorate,decoration={zigzag,amplitude=2pt,segment length=4pt}] 
        (-3,-3) .. controls (-1.5,-2.5) and (1.5,-2.5) .. (3,-3); 

    \draw[thick] (-3,3) -- (3,-3);
    \draw[thick] (-3,-3) -- (3,3);

    \draw[thick,blue] (-2.4,2.9) -- (3,-2.5);

    \draw[thick,red] (1.1,-0.6) -- (3.0,1.4);
    \draw[thick,red,dashed] (1.89,-1.28) -- (3.0,-0.1);
    \draw[thick,red] (0.55,-2.58) -- (1.89,-1.28);

    \node[rotate=45] at (2.0,1.7) {$u \, (v = 0)$};
    \node[rotate=-45] at (-2.0,1.7) {$v \, (u = 0)$};

    \draw[<->,black] (2.1,-1.1) -- (1.38,-0.36);
    \node[rotate=30,black] at (2.1,-0.4) {$\bar{h}(x)$};
\end{tikzpicture}
    \caption{The Penrose diagram (corresponding to a two-sided black hole geometry) illustrates the impact of a shock wave generated near the horizon of the black hole. This shock wave, shown as a blue line, originates from the perturbation created by the particle \( W \), which is inserted at an early enough time. Without the shock wave, the \( V \) particle would have emerged from the past interior region and reached the boundary at \( t = 0 \). However, the presence of the shock wave alters its trajectory, introducing a shift parametrized by \( \bar{h}(x) \). This shift results in a delay in the appearance of the operator \( V \) at the boundary. When the \( W \) perturbation is inserted sufficiently early ( \( |t| \gtrsim |t_*| \) ) the trajectory of \( V \) is displaced so significantly that it falls into the future interior, preventing it from reaching the boundary at all.}
    \label{fig:OTOC}
\end{figure}

In the AdS/CFT correspondence, an operator $\hat{Q}$ in the boundary corresponds to a field $\phi(x,t)$ in the bulk. Suppose we act with an operator $W(0,-t)$ on $\ket{TFD}$ in the right boundary. In the two-sided black hole geometry, this corresponds to a particle coming out from the past interior, reaching the boundary at time $-t$, and then falling toward the future interior. The energy of the particle falling into the black hole increases exponentially, and it depends on the temperature of the black hole
\begin{equation}
    \label{eqn:expeng}
    E = E_{o}e^{\frac{2\pi}{\beta}t}\,,
\end{equation}
where $E_{o}$ is the energy of the particle when it is near the boundary. Since the energy of the particle gets exponentially blueshifted, a sufficiently earlier perturbation in the boundary, $W(0,-t)$, leads to a nontrivial modification of the bulk geometry. The large energy of the particle leads to a backreaction in the geometry, which simply corresponds to a shock wave geometry \cite{Roberts:2014isa}. The energy distribution of this perturbation is compressed along the $u$ direction and stretched along the $v$ direction. This is illustrated in Figure~\ref{fig:OTOC}. Thus, for sufficiently large time $|t|$, the perturbation gets localized along the horizon $u = 0$. The change in the stress-energy tensor sources this localized shock wave:
\begin{equation}
\label{eqn:Tshock}
   \delta T_{uu}^{shock} = E_{o}e^{\frac{2\pi}{\beta}t}\delta(u)a(x)\,,
\end{equation}
where $a(x)$ is some function to be determined. The effect of this shock, localized along the horizon $u=0$, is that the particle coming from the past interior suffers a shift in the trajectory (refer to Figure~\ref{fig:OTOC}), which is given as
\begin{equation}
    \label{eqn:shift}
    v \rightarrow v + \Theta(u) \bar{h}(x), \hspace{0.5cm} u \rightarrow u\,,
\end{equation}
where the step function $\Theta(u)$ accounts for the fact that the causal future of the particle is affected after it encounters the horizon $u=0$, and thus, there is no effect on the causal past. $\bar{h}(x)$ is a function that characterizes the amount of shift, and it is to be determined by the Einstein equation of motion.\footnote{The shift function $\bar{h}(x)$ should not be confused with function $h(z)$ used in the spacetime metric in Eq.~(\ref{eqn:generalmetric}).}

The function $\bar{h}(x)$ is closely related to the commutator \eqref{eqn:expcommutator}. To see why that is the case, consider again the overlap of the two states \eqref{eqn:twostates} in the holographic setup with $\ket{\beta}$ being the thermofield double state $\ket{\text{TFD}}$. In the two-sided black hole geometry, acting with an operator $V$ or $W$ on the right boundary corresponds to a perturbation (which we will refer to as the $V$ particle or $W$ particle, respectively) that emerges from the past interior and reaches the boundary of the right exterior. Here, the bulk picture of $\ket{\psi_{1}}$ can be understood as follows: Initially, the system is prepared in such a way that the $V$ particle from the past interior reaches the boundary at $t=0$ [i.e., $V(x,0)\ket{\text{TFD}}$]. Then the state $V(x,0)\ket{\text{TFD}}$ is evolved backward in time, the perturbation operator $W$ is acted at time $-t$, and then the overall state is evolved forward in time again. If the operator $W$ is inserted at a sufficiently earlier time (i.e., $ t \sim t_{*}$), the $W$ particle reaches and creates a shock wave at the horizon ($u=0$). This causes the $V$ particle to encounter the shock wave at $u = 0$ at a later time and causes a shift in its trajectory parametrized by $\bar{h}(x)$. Because of this shift in the trajectory of the $V$ particle, the $V$ operator does not appear on the boundary at $t=0$, but its appearance gets time delayed. If the $W$ operator is acted at a sufficiently earlier time, the shock wave may become strong enough to shift the trajectory of the $V$ particle such that it gets engulfed into the future interior and, correspondingly, the $V$ operator does not rematerialize on the boundary at any time. Similarly, the state $\ket{\psi_2}$ can be understood as follows: The operator $W$ is acted on $\ket{\text{TFD}}$ at a sufficiently early time $-t$, and the $W$ particle creates a shock wave at the horizon. Then, the $V$ particle suffers a shift in the trajectory due to the shock wave at $u=0$ in such a way that, after suffering the shift, it reaches the boundary at $t=0$. Thus, the overlap between the two states, or the OTOC, decreases as the time delay in the appearance of $V$ (in the first case) increases. This implies that, as $\bar{h}(x)$ increases (and thereby the shift), the overlap or the OTOC decreases. This clearly suggests a relation between $\bar{h}(x)$ and the commutator \eqref{eqn:expcommutator}, and that $\bar{h}(x)$ should also attain an exponential form. Indeed, as we shall see, by solving for $\bar{h}(x)$ from the Einstein equation, it attains an exponential form like $\eqref{eqn:expcommutator}$. By comparing $\bar{h}(x)$ and $\eqref{eqn:expcommutator}$, we can read off $v_B$, $\lambda_L$, and $t_*$.

One can also physically understand why $\bar{h}$ should depend on the coordinate $x$. As noted before, after the scrambling time $t_*$, the information about the perturbation starts scrambling among the local degrees of freedom at a rate $v_B$, with the size of the operator given by $|x| \leq v_B (t - t_*)$. For a given time difference $t-t_*$ and rate $v_B$, the effect of the perturbation operator $W$ on the probe operator $V$ decreases as the spatial separation $x$ increases. This implies that the commutator decreases with increasing $x$, and thereby $\bar{h}$ decreases too. Hence, clearly, $\bar{h}$ is a function of $x$. 

With this physical picture in mind, we now proceed to explicitly compute the function $\bar{h}(x)$.
The shift in the coordinates \eqref{eqn:shift} results in the following modification in the metric:
\begin{equation}
\label{eqn:modifiedmetric}
    ds^{2} = 2A(u,v)dudv + B(u,v)dx^i dx^i - 2A(u,v)\bar{h}(x)\delta(u)du^{2}\,.
\end{equation}
The total stress tensor is now given as
\begin{equation}
    \label{eqn:totalT}
    T = T_{o} + \delta T^{shock}\,,
\end{equation}
where $T_{o}$ is the initial unperturbed stress tensor and the only nonzero component of $\delta T^{shock}$ is given by \eqref{eqn:Tshock}. Substituting \eqref{eqn:Tshock} and \eqref{eqn:modifiedmetric} in the Einstein equation and solving for the $uu$ component, we arrive at the following equation:
\begin{equation}
    \label{eqn:odeforh}
    \left(\partial_{i}\partial_{i} - \frac{d-1}{2}\frac{\partial_{u}\partial_{v}B}{A}\right) h(x)\delta(u) = \frac{8 \pi G_{N} E_{o} B}{A} e^{\frac{2\pi t}{\beta}}\delta(u)a(x)\,.
\end{equation}
Here, $d$ is the number of boundary spacetime dimensions. For large $x$ ($|x| >> 1$), the function $a(x)$ can be replaced by the Dirac $\delta$ delta function, as the solution depends only on the integral of $a(x)$. Solving for large $x$, one gets the following solution for $\bar{h}(x)$:
\begin{equation}
    \label{eqn:hxt}
    \bar{h}(x) \sim \frac{e^{\frac{2\pi}{\beta}(t-t_{*})- \chi |x|}}{|x|^{\frac{d-2}{2}}}\,,
\end{equation}
with $\chi$ and $t_{*}$ given as
\begin{equation}
    \label{eqn:chi}
    \chi = \sqrt{\frac{d-1}{2}\left(\frac{\partial_{u}\partial_{v}B(0)}{A(0)}\right)}\,,
\end{equation}
\begin{equation}
    \label{eqn:tstar}
    t_{*} = \frac{2 \pi}{\beta} \log\left(\frac{A(0)}{8 \pi G_{N}E_{o}B(0)}\right)\,.
\end{equation}
Here, $\partial_{u}\partial_{v}B(0)$ denotes the double derivative of $B$ evaluated at the horizon and $\{ A(0), B(0)\}$ denote the functions $\{A, B\}$ evaluated at the horizon. Comparing \eqref{eqn:hxt} with \eqref{eqn:expcommutator}, we can read off the butterfly velocity
\begin{equation}
    \label{eqn:VBOTOC}
    v_{B} = \frac{2\pi}{\beta \chi}\,.
\end{equation}

\subsection{Pole skipping}
\label{sec:poleskipping}
This section discusses the pole-skipping (PS) phenomenon to compute the butterfly velocity. The PS phenomenon occurs when lines of poles and zeros of the retarded Green's function intersect, leading to specific values of frequency ($\omega_*$) and momentum ($q_*$) at which the Green's function does not yield a unique value. In \cite{Grozdanov:2017ajz, Blake:2017ris}, it was demonstrated that these special points can be obtained by a hydrodynamical effective field theory description and are given by 
\begin{equation}
    \label{eqn:specialpoint}
    \omega_{*}=i \lambda_{L}, \hspace{0.3cm} q_{*}=i \frac{\lambda_{L}}{v_{B}}\,,
\end{equation}
where, as usual, $\lambda_L$ and $v_{B}$ are the Lyapunov exponent and the butterfly velocity. The above relations can be realized from the two-point energy density correlation function, which shows a special behavior at points \eqref{eqn:specialpoint} in Fourier space. Essentially, the retarded Green's function of the energy density takes the form 
\begin{equation}
    \label{eqn:GreensFunction}
    G^{R}_{T^{00}T^{00}}=\frac{B(\omega,q)}{A(\omega,q)}\,.
\end{equation}
The pole and the zero lines of $G^{R}_{T^{00}T^{00}}$ are given by $A(\omega,q)=0$  and $B(\omega,q)=0$, respectively. At the point of intersection of these two special lines, we have
\begin{equation}
    \label{eqn:GreensBandAatPSpoint}
B(\omega_{*},q_{*})=A(\omega_{*},q_{*})=0\,,
\end{equation}
which allows us to define the pole-skipping point $(\omega_{*},q_{*})$. Therefore,  by computing the special points $\omega_*$ and $q_*$, one can obtain the chaotic information of the system from Eq.~(\ref{eqn:specialpoint}). Note that the pole and zero lines of the Green's function can meet at multiple points in the complex momentum space $(\omega,q)$. Our interest here is not in computing all those pole-skipping points. Our objective is only to compute the butterfly velocity using \eqref{eqn:specialpoint} from the pole-skipping point, which is related to the prominent chaotic behavior of the system.

Note that $G^{R}_{T^{00}T^{00}}$ has an indeterminate form at $(\omega_{*},q_{*})$.
Near the point $(\omega_{*},q_{*})$, the Green's function can be evaluated by the following limit:
\begin{equation}
    \label{eqn:GreensnearPSpoint}
    \begin{split}
    G^{R}_{T^{00}T^{00}}&=\lim_{\Delta\omega \to 0,\Delta q \to 0}   \frac{B(\omega_{*}+\Delta\omega ,q_{*}+\Delta q)}{A(\omega_{*}+\Delta\omega ,q_{*}+\Delta q)}\\
    &=\lim_{\epsilon \to 0}   \frac{B(\omega_{*}+\epsilon\delta\omega ,q_{*}+\epsilon\delta q)}{A(\omega_{*}+\epsilon\delta\omega ,q_{*}+\epsilon\delta q)}
    = \frac{(\partial_{\omega}B)_{*}+\frac{\delta q}{\delta\omega}(\partial_{q}B)_{*}}{(\partial_{\omega}A)_{*}+\frac{\delta q}{\delta\omega}(\partial_{q}A)_{*}}\,.
    \end{split}
\end{equation}
Since the limit depends on the slope $\delta q/\delta\omega$ of the curve at the point $(\omega_{*},q_{*})$, via which the pole-skipping point is approached, $G^{R}_{T^{00}T^{00}}$ is not uniquely defined at the point $(\omega_{*},q_{*})$.  Thus, the Green function's is multivalued at $(\omega_{*},q_{*})$ \cite{Blake:2018leo, Blake:2019otz, Natsuume:2019sfp}.

The main quantities of interest to us are the pole-skipping points $\omega_{*}$ and $q_{*}$. These can be computed from the near-horizon expansion of the linearized Einstein equations. In particular, in the gauge/gravity duality formalism, the correlation function $G^{R}_{T^{00}T^{00}}(\omega,q)$ is related to the gravitational perturbation equation of the sound channel, subject to the ingoing boundary conditions at the horizon. To implement the ingoing boundary condition and to simplify the computation, it is generally convenient to use the ingoing Eddington-Finkelstein coordinates $(v,z,x^i)$,  where $v=t+z_{*}$.  In these coordinates, the background metric \eqref{eqn:generalmetric} takes the form
\begin{equation}
    \label{eqn:EFgeneralmetric}
    ds^{2} = - h(z)g(z)dv^{2} +2\sqrt{h(z)f(z)}dvdz+ k(z)dx^{i}dx^{i}\,,
\end{equation}
with $z_*$ given by (\ref{eqn:tortoise}). Next, we  perturb the background metric
\begin{eqnarray}
& & g_{\alpha\beta} \rightarrow g_{\alpha\beta}+\delta g_{\alpha\beta}(z)e^{-i\omega v+iqx}\,,
\label{metricpertb}
\end{eqnarray}
and perform the Fourier transformation of the perturbations. For simplicity, we assume that the metric (and matter) perturbations propagate solely along the $x^1 \equiv x$ direction. By imposing the radial gauge condition $\delta g_{z \alpha} = 0$, it can be shown that the only perturbation fields that couple in the sound channel are $\delta g_{vv},\delta g_{vx},\delta g_{xx}$, and $\delta \psi_{i}$, where $\delta \psi_{i}$ represents fluctuation in the matter fields collectively.

As previously stated, the appropriate Green's function is determined by the perturbed equations, which are regular at the black hole horizon and subject to the ingoing boundary condition. For this purpose, we consider the following near-horizon expansion of the metric and the matter perturbations:
\begin{equation}
\label{eqn:nearhorizonexpansion} 
X(z) = \sum_{n=0}^{\infty} X^{(n)}(z_h) \frac{(z - z_h)^n}{n!}\,,
\end{equation}
where $X$ collectively represents the perturbations of metric and matter fields, and the superscript $n$ denotes the $n{\text{th}}$ order of derivative with respect to $z$. To proceed, we further need to define weight. Following \cite{Wang:2022mcq, Ning:2023ggs}, it is defined for any tensor component as the number of upper $z$ indices minus the number of upper $v$ indices, where lower $v$ indices are considered as upper $z$ indices and vice versa.

In \cite{Wang:2022mcq}, it was established that, for any arbitrary diffeomorphism invariant Einstein gravity with matter fields, the first instance of pole-skipping points happens at frequency $\omega=i(p_0-1)2\pi T$, where $p_0$ is the highest weight. For metric perturbations like in Eq.~(\ref{metricpertb}), the highest weight is $p_0=2$. Accordingly, from  Eq.~\eqref{eqn:specialpoint}, it is easy to see that the first pole-skipping point directly corresponds to maximal Lyapunov exponent and chaos bound $\lambda_{L}=2\pi T$. Note that the butterfly velocity obtained from the first pole-skipping point for $p_0=2$ matches the one obtained from the OTOC \cite{Wang:2022mcq}. It suggests that the first pole-skipping point can be obtained from the Einstein field equation of the highest weighted perturbation field.

Among the above-mentioned perturbation fields in the sound channel, the highest weight is $2$, which is associated with $\delta g_{vv}$. The corresponding equation of motion is the linearized Einstein field equation $\delta E_{vv}=\delta T_{vv}$. For matter fields relevant for our purposes (see the next sections), the stress tensor perturbation $\delta T_{vv}(z_h)$ may not vanish. However, $\delta T_{v}^{z}$ does vanish for matter fields that are regular at the horizon \cite{Jeong:2021zhz, Baishya:2024gih}. Keeping this in mind, we set $\delta T_{v}^{z}=0$ from here on to evaluate the pole-skipping points relevant to chaos.

Substituting the near-horizon expansion \eqref{eqn:nearhorizonexpansion} of the various metric and matter perturbations into $\delta E_{v}^{z}=\delta T_{v}^{z}$ and comparing the coefficients of $(z-z_h)^n$, we obtain a set of equations (say, $\boldsymbol{\mathcal{S}}_n=0$) that puts a nontrivial constraint on $\delta g_{vv},\delta g_{vx}, \delta g_{xx}$, and $\delta \psi_{i}$. The equation $\boldsymbol{\mathcal{S}}_0=0$ takes the following form:
\begin{equation}
\label{eqn:generalperturbEinsteinsequation} 
\begin{split}
\left(2q\delta g_{vx}^{(0)}(z_{h}) + \omega\delta g_{xx}^{(0)}(z_{h})\right) &\left(2\omega  - i g'(z_{h})\sqrt{\frac{h(z_{h})}{f(z_{h})}}\right) + \left(2q^{2} - (d-1) \frac{i\omega k'(z_{h})}{\sqrt{h(z_{h})f(z_{h})}}\right) \delta g_{vv}^{(0)}(z_{h})
\\ &= 4k(z_{h})\left(\delta T_{vv}(z_{h}) - \frac{1}{\sqrt{h(z_{h})f(z_{h})}}T_{vz}(z_{h})\delta g_{vv}^{(0)}(z_{h})\right)\,,
\end{split}
\end{equation}
where $\delta T_{vv}(z_{h})$ in the RHS of the above equation is the $vv$ component of the perturbed bulk stress-energy tensor, which only depends on the matter content coupled to the background metric, i.e., the precise form of matter Lagrangian $\mathcal{L}_{M}$. In the next section, we consider various forms of matter action, relevant for holographic QCD, and demonstrate that $\delta(T_{v}^{z})$ equates to zero for these models. For now, let us assume that $\delta(T_{v}^{z})$ indeed vanishes. This leads to 
\begin{equation}
\label{eqn:RHSgeneralperturbEE} 
\begin{split}
\delta T_{vv}(z_{h}) - \frac{1}{\sqrt{h(z_{h})f(z_{h})}}T_{vz}(z_{h})\delta g_{vv}^{(0)}(z_{h})=0 \,.
\end{split}
\end{equation}
Therefore, Eq.~\eqref{eqn:generalperturbEinsteinsequation} simplifies to
\begin{equation}
\label{eqn:generalperturbEEwithTterm} 
\begin{split}
\left(2q\delta g_{vx}^{(0)}(z_{h}) + \omega\delta g_{xx}^{(0)}(z_{h})\right) \left(2\omega  - 4i\pi T\right) + \left(2q^{2} - (d-1)i\omega \frac{k'(z_{h})}{\sqrt{h(z_{h})f(z_{h})}}\right) \delta g_{vv}^{(0)}(z_{h})  
= 0\,,
\end{split}
\end{equation}
where we have used Eq.~\eqref{eqn:tempbh}. For any general $\omega$ and $q$, the above equation provides a nontrivial constraint relating $\delta g_{vx}^{(0)}(z_{h})$, $\delta g_{xx}^{(0)}(z_{h})$, and $\delta g_{vv}^{(0)}(z_{h})$. When $\omega=2i\pi T$, the above equation puts no constraint on $\delta g_{vx}^{(0)}(z_{h})$ and $\delta g_{xx}^{(0)}(z_{h})$ and takes the form
\begin{equation}
\label{eqn:perturbEEwithonlygvv} 
\begin{split}
\left(q^{2} + (d-1)\pi T \frac{k'(z_{h})}{\sqrt{h(z_{h})f(z_{h})}}\right) \delta g_{vv}^{(0)}(z_{h})  
= 0 \,.
\end{split}
\end{equation}
If the coefficient of $\delta g_{vv}^{(0)}(z_{h})$ also vanishes, the equation holds trivially. Thus, Eq.~\eqref{eqn:generalperturbEEwithTterm} becomes trivial for the following values of $\omega$ and $q$:
\begin{equation}
    \label{eqn:TtermPSpointcomputated}
    \omega=2 i \pi T; \hspace{0.3cm} q=i\sqrt{(d-1)\pi T \frac{k'(z_{h})}{\sqrt{h(z_{h})f(z_{h})}}}\,.
\end{equation}
For the above values of $\omega$ and $q$, there is one less equation to solve for $\delta g_{vx}^{(0)}(z_{h})$, $\delta g_{xx}^{(0)}(z_{h})$, and $\delta g_{vv}^{(0)}(z_{h})$ than for any other general value of $\omega$ and $q$. As a result, there is an extra linearly independent ingoing solution (extra ingoing mode) to the linearized Einstein equation at the horizon. When we choose the ingoing mode at the above value of $\omega$ and $q$, unlike at generic $\omega$ and $q$, the ratio $B/A$  does not have a definite value, and the energy density retarded Green's function becomes infinitely multivalued, as having an extra ingoing mode leads to having no constraints on $A$ and $B$. More details on this can be found in  \cite{Blake:2018leo}. Therefore, $\omega$ and $q$ given by \eqref{eqn:TtermPSpointcomputated} constitute the pole-skipping point $(\omega_{*},q_{*})$ in the Fourier space.

Using the pole-skipping points, we arrive at the following expressions for $\lambda_{L}$ and $v_{B}$:
\begin{equation}
    \label{eqn:generalPSlambdaandvb}
    \lambda_{L}=2\pi T\,,~~~ \hspace{0.3cm} v_{B}^{2} =\frac{\omega_{*}^2}{q_{*}^2}= \frac{h_{o}g_{o}'}{(d-1)k_{o}'}\,.
\end{equation}
Notice that this is the same expression that we derived in Section \ref{sec:EW} using the entanglement wedge method. Accordingly, we will get the same thermal- and chemical-potential-dependent spectrum for the butterfly velocity from these two methods. Thus, in the next section, where we explicitly compute $v_{B}$ for various holographic QCD models, and all we have to do is to check whether the matter action of the models satisfies $\delta T_{v}^{z} = 0$ to see if the pole-skipping method yields the same result as the other methods. 

\section{Butterfly velocity in 1RCBH and 2RCBH holographic models}
\label{subsec:RCBH}
\subsection{Background}
In this section, we consider two top-down holographic models that describe a strongly coupled quantum field theory at finite density and temperature, namely, the 1RCBH and 2RCBH models \cite{DeWolfe:2011ts, DeWolfe:2012uv}. The Einstein-Maxwell-dilaton (EMD) action in five dimensions for both of these models is given by
\begin{equation}
    \label{eqn:EMDaction}
    S = \frac{1}{16 \pi G_5} \int d^5x \sqrt{-g} \left[ R - \frac{ \tilde{f}(\phi)}{4} F_{MN}F^{MN} - \frac{1}{2} (\partial_M \phi)^2 - V(\phi) \right]\,,
\end{equation}
where $R$ is the Ricci scalar associated with the five-dimensional bulk metric $g_{MN}$, $V(\phi)$ is the potential of the dilaton field $\phi$, $F_{MN}=\partial_M \mathcal{A}_N -\partial_N \mathcal{A}_M$ is the Maxwell field strength, and $\tilde{f}(\phi)$ represents the coupling between the Maxwell and dilaton fields. $G_5$ is Newton's constant in five dimensions.

Both 1RCBH and 2RCBH models are cases of a more general top-down holographic model called the STU model \cite{Behrndt:1998jd,Cvetic:1999ne}. The STU gravity model describes five-dimensional black brane solutions containing three conserved Abelian $R$ charges. The 1RCBH model is obtained when two $R$ charges are set to zero, whereas the 2RCBH model is obtained by setting one $R$ charge to zero and further identifying the remaining two $R$ charges. Therefore, in each case, there is only one nontrivial charge and hence the chemical potential. Essentially, the dual field theory configurations of the 1RCBH model should be thought of as states in $\mathcal{N}=4$ super-Yang-Mills (SYM) theory with a temperature and a chemical potential for a $U(1)$ subgroup of the $SO(6)$ $R$ symmetry. Similarly, the 2RCBH model corresponds to a two-charge $\mathcal{N}=4$ solution, which has equal charges for two $U(1)$ gauge fields inside the $SO(6)$ $R$ symmetry. The black hole solutions of these two models (discussed below) correspond to a strongly coupled charged SYM plasma in the dual boundary side.

The two models represent different realizations of the strongly coupled $\mathcal{N}=4$ SYM plasma at finite temperature and charge density. Since these models are analytically solvable, and geometric and thermodynamic properties are known analytically, they have been greatly used in holographic literature. Both exhibit several features in common with QCD-like black holes, i.e., models where the dilaton potential and gauge-kinetic coupling are tuned to reproduce lattice QCD results \cite{DeWolfe:2011ts}. The 1RCBH model features a critical point and has been used to study the QCD phase diagram and thermodynamics \cite{DeWolfe:2011ts}. Various transport phenomena relevant to QCD, such as conductivity and diffusion, have also been investigated within this model \cite{DeWolfe:2011ts}. In addition to thermodynamics and transport, several quantum information observables \cite{Ebrahim:2018uky} \cite{Ebrahim:2020qif} and chaotic properties~\cite{Amrahi:2023xso} \cite{Karan:2023hfk} have been explored. In this work, we provide a more comprehensive analysis of the model's chaotic behavior. 
On the other hand, the 2RCBH model does not exhibit any critical point. Nevertheless, it has been employed to study a range of transport phenomena in QCD \cite{DeWolfe:2012uv}, nonequilibrium homogeneous isotropization dynamics \cite{deOliveira:2025lhx}, nonhydrodynamic quasinormal modes and late-time equilibration of the plasma \cite{deOliveira:2024bgh}. Moreover, this model captures non-Fermi liquid behavior, exhibiting nested Fermi surfaces with non-quasi-particle excitations, phenomena which could be potentially relevant to dense QCD matter \cite{DeWolfe:2012uv}. To the best of our knowledge, the chaotic properties of the 2RCBH model have not been previously explored. Our work is, therefore, the first to present a detailed analysis of chaos in the 2RCBH model.

Despite both originating from the STU model and describing strongly interacting thermal plasma at finite $R$-charge densities, the 2RCBH and 1RCBH plasmas exhibit distinct behaviors. While both reduce to a thermal SYM plasma at zero $R$-charge density, the finite $R$-charge chemical potential leads to differing phase diagrams. In the 2RCBH model, the ratio of chemical to temperature $\mu/T$  ranges from zero to infinity without a phase transition, while the 1RCBH model exhibits a critical point and $\mu/T$ reaches a maximum at the critical point.

Essentially, 1RCBH and 2RCBH gravity models correspond to the EMD action with the following dilaton potentials $V(\phi)$ and coupling functions $\tilde{f}(\phi)$:
\begin{eqnarray}
& & V_1(\phi) = V_2(\phi) = - \frac{1}{L^2}\left( 8 e^{\phi/\sqrt{6}} + 4e^{-\sqrt{2/3}\phi} \right)\,, \nonumber\\
& & \tilde{f}_1(\phi) = e^{-2\sqrt{2/3}\phi},~~~\tilde{f}_2(\phi) = e^{\sqrt{2/3}\phi}\,,
\end{eqnarray}
where subscripts $1$ and $2$ are used for 1RCBH and 2RCBH models, respectively. Here $L$ is the AdS radius, which we will set to $1$ from now on for convenience. For the following Ans\"atze for the metric, gauge field, and dilaton field:
\begin{eqnarray}
\label{eqn:gubsermetric} 
& & ds^{2} = e^{2A(z)}\left[-H(z)dt^{2}+dx_{1}^{2}+dx_{2}^{2}+dx_{3}^{2}\right] + \frac{e^{2B(z)}}{H(z)}dz^{2}\,, \nonumber \\
& & \mathcal{A}_M=\Phi(z)\delta_{M}^{0},~~ \phi=\phi(z)\,.
\end{eqnarray}
The gravity solution of the 1RCBH EMD model is given by
\begin{eqnarray}
 & & A_{1}(z) = \ln(z) + \frac{1}{6} \ln\left(1+\frac{Q_{1}^{2}}{z^{2}}\right)\,,~~~ B_{1}(z) = - \ln(z) - \frac{1}{3}\ln\left(1+\frac{Q_{1}^{2}}{z^{2}}\right)\,, \nonumber\\
& & H_{1}(z) = 1 - \frac{z_h^{2}(z_h^{2} + Q_{1}^{2})}{z^{2}(z^{2}+Q_{1}^{2})}\,,~~~ \Phi_1(z) = \frac{z_h Q_1}{\sqrt{z_{h}^{2}+Q_{1}^{2}}}\left(1- \frac{z_{h}^{2}+Q_{1}^{2}}{z^{2}+Q_{1}^{2}} \right)\,,  \nonumber\\
& & \phi_1(z)=-\sqrt{\frac{2}{3}}\ln\left( 1+ \frac{Q_{1}^{2}}{z^2}  \right)\,.
\label{1RCBHmetric}
\end{eqnarray}
Similarly, the gravity solution of the 2RCBH EMD model is given by
\begin{eqnarray}
 & & A_{2}(z) = \ln(z) + \frac{1}{3} \ln\left(1+\frac{Q_{2}^{2}}{z^{2}}\right)\,,~~~ B_{2}(z) = - \ln(z) - \frac{2}{3}\ln\left(1+\frac{Q_{2}^{2}}{z^{2}}\right)\,, \nonumber\\
& & H_{2}(z) = 1 - \frac{(z_h^{2} + Q_{2}^{2})^2}{(z^{2}+Q_{2}^{2})^2}\,,~~~ \Phi_2(z) = \sqrt{2}Q_2 \left(1- \frac{z_{h}^{2}+Q_{2}^{2}}{z^{2}+Q_{2}^{2}} \right)\,,  \nonumber\\
& & \phi_2(z)=\sqrt{\frac{2}{3}}\ln\left( 1+ \frac{Q_{2}^{2}}{z^2}  \right)\,.
\label{2RCBHmetric}
\end{eqnarray}
In the above solutions, $z_h$ is the black hole horizon, determined by the largest real root of the blackening function $H_i(z)$, and $Q_i$ are the black hole charges. The chemical potential of the dual boundary theory $\mu_i$, corresponding to the leading-order term of the zeroth component of the gauge field at the asymptotic boundary, is related to the black hole charge in the following way: 
\begin{eqnarray}
 & & \mu_1 = \frac{z_h Q_1}{\sqrt{z_{h}^{2}+Q_{1}^{2}}},~~~~~\mu_2 = \sqrt{2} Q_2\,.
\label{chemicalpotentialgubser}
\end{eqnarray}
Similarly, the temperature and entropy of 1RCBH and 2RCBH black brane solutions are given by
\begin{eqnarray}
 & & T_1 = \frac{2z_{h}^{2}+Q_{1}^{2}}{2 \pi\sqrt{z_{h}^{2}+Q_{1}^{2}}},~~~~~ T_2 =  \frac{z_h}{\pi}\,, \nonumber\\
& & S_1 = \frac{z_{h}^{3}\sqrt{1+\frac{Q_{1}^{2}}{z_{h}^{2}}}V_3}{4G_5},~~~~S_2 = \frac{z_{h}^{3}\left(1+\frac{Q_{1}^{2}}{z_{h}^{2}}\right) V_3}{4G_5}\,,
\label{tempgubser}
\end{eqnarray}
where $V_3$ is the unit volume of the three-dimensional boundary space. Note that, in the above solutions, $z$ corresponds to the usual holographic radial coordinate and it runs from $z=z_h$ (horizon radius) to $z=\infty$ (asymptotic boundary).

Using the above thermodynamic observables, one can analyze the thermodynamic behavior and stability of 1RCBH and 2RCBH black brane solutions. It turns out that there are two black hole branches - large and small - in the 1RCBH solution. The large black hole branch (large $z_h$ solution) is thermodynamically stable and exhibits a positive specific heat, whereas the small black hole branch (large $z_h$ solution) is thermodynamically unstable and exhibits negative specific heat. There also appears a critical point $\mu_1=\pi T_1/\sqrt{2}$ at which both branches merge and lead to diverging specific heat. Accordingly, the range of values of $\mu_1/T_1$ probed by the 1RCBH model is restricted to $\mu_1/T_1 \in [0,\pi/\sqrt{2}]$. As discussed in \cite{Finazzo:2016mhm}, $\mu/T = \pi/\sqrt{2}$ is a critical point of second-order phase transition. Note that this phase transition is not the usual Hawking/Page phase transition. On the other hand, there is only one black hole branch in the 2RCBH model, which is always thermodynamically stable. Moreover, the allowed range of $\mu_2/T_2$ is unrestricted in the 2RCBH model; i.e., it probes values of  $\mu_2/T_2$ ranging from zero to infinity. Below, in the computation of the butterfly velocity, we will work in the thermodynamically stable phases and restrict ourselves to only those values of the chemical potential and temperature that are allowed in 1RCBH and 2RCBH black brane solutions.  

\subsection{Calculation of $v_{B}$ using the entanglement wedge method}
We first compute the butterfly velocity for the 1RCBH model. Here, we concentrate only on the large black hole solution, corresponding to a stable thermal state. Comparing the 1RCBH  metric \eqref{eqn:gubsermetric} with the general metric \eqref{eqn:generalmetric}, we have the following for the 1RCBH model\footnote{Here $f(z)$ should not be confused with the coupling functions $\tilde{f}_1(\phi)$ and $\tilde{f}_2(\phi)$.}: 
\begin{equation}
\begin{split}
    \label{eqn:metricfunctions1RCBH}
    &g(z) = H_1(z) = 1 - \frac{z_h^{2}(z_h^{2} + Q_{1}^{2})}{z^{2}(z^{2}+Q_{1}^{2})}, \hspace{0.3cm} \\ &h(z) = k(z) = e^{2A_{1}(z)} = z^{2}\left(1+\frac{Q_{1}^{2}}{z^{2}}\right)^{\frac{1}{3}}, \hspace{0.3cm} \\ &f(z) = e^{2B_{1}(z)} =\frac{1}{z^{2}\left(1+\frac{Q_{1}^{2}}{z^{2}}\right)^{\frac{2}{3}}}.
\end{split}
\end{equation}
Computing the metric functions and their derivatives at the horizon and substituting in \eqref{eqn:vbfinal}, we get the following expression for the butterfly velocity:
\begin{equation}
    \label{eqn:vb1RCBH}
    v_{B}^{2} =  \frac{Q_{1}^{2} + 2 z_h^{2}}{ 2Q_{1}^{2}+ 3z_h^{2}}\,.
\end{equation}
We can express $v_B$ in terms of $\mu_1$ by substituting $Q_1$ from \eqref{chemicalpotentialgubser} into \eqref{eqn:vb1RCBH},
\begin{equation}
    \label{eqn:vb1RCBHEW}
    v_{B}^{2} = \frac{2}{3}\left(\frac{z_h^{2}-\mu_{1}^{2}/2}{z_h^{2}-\mu_{1}^{2}/3}\right)\,.
\end{equation}
In this work, we wish to study the behavior of the butterfly velocity with respect to the chemical potential and the temperature. In order to achieve this for the 1RCBH model, we have to replace $z_h$ in \eqref{eqn:vb1RCBHEW} in terms of $\mu_1$ and $T_1$. This can be done by first solving for $Q_1$ in terms of $T_1$ and $z_h$ from \eqref{tempgubser}, substituting it in \eqref{chemicalpotentialgubser}, and finally solving for $z_h$ in terms of $\mu_1$ and $T_1$. The $z_h$ corresponding to the thermodynamically stable branch is obtained as
\begin{eqnarray}
    \label{eqn:zhmuT1RCBH}
    z_h (\mu_1,T_1) = \frac{\sqrt{\pi^2T_{1}^{2} + \mu_{1}^{2} + \pi T_1 \sqrt{\pi^2T_{1}^{2}-2\mu_{1}^{2}}}}{\sqrt{2}}\,.
\end{eqnarray}
Substituting \eqref{eqn:zhmuT1RCBH} into \eqref{eqn:vb1RCBHEW}, we get the following expression of $v_B$ in terms of $\mu_1$ and $T_1$ for the 1RCBH model:
\begin{equation}
\label{eqn:vb1RCBHmuT}
    v_B^2 = \frac{2\pi T_1(\pi T_1 + \sqrt{\pi^2 T_1^2 - 2\mu_1^2})}{3 \pi^2T_1^2 + \mu_1^2 + 3\pi T_1\sqrt{\pi^2T_1^2 - 2\mu_1^2}}\,.
\end{equation}

We can similarly compute $v_B$ for the 2RCBH model. The metric functions for the 2RCBH model are given as
\begin{equation}
\begin{split}
    \label{eqn:metricfunctions2RCBH}
    &g(z) = H_2(z) = 1 - \frac{(z_h^{2} + Q_{2}^{2})^{2}}{(z^{2}+Q_{2}^{2})^{2}}\,, \\ &h(z) = k(z) = e^{2A_{2}(z)} = z^{2}\left(1+\frac{Q_{2}^{2}}{z^{2}}\right)^{\frac{2}{3}}\,, \\
    & f(z) = e^{2B_{2}(z)} = \frac{1}{z^{2}\left(1+\frac{Q_{2}^{2}}{z^{2}}\right)^{\frac{4}{3}}}.
\end{split}    
\end{equation}
Substituting the near-horizon values of these metric functions into \eqref{eqn:vbfinal} and further expressing $Q_2$ in terms of $\mu_2$ from \eqref{chemicalpotentialgubser}, we get the following expression for the butterfly velocity for the 2RCBH model:
\begin{equation}
    \label{eqn:vb2RCBHfinal}
    v_{B}^{2} = \frac{2}{3}\left(\frac{z_h^{2}}{z_h^{2} + \mu_{2}^{2}/6}\right)\,.
\end{equation}
For the 2RCBH model, $z_h(\mu_2,T_2)$ is simply $z_h (\mu_2, T_2) = \pi T_2$. Substituting $z_h(\mu_2, T_2)$ into \eqref{eqn:vb2RCBHfinal}, we can get $v_B$ in terms of $\mu_2$ and $T_2$ for the 2RCBH model. 

\begin{figure}[h]
    \centering
    \begin{subfigure}[b]{0.4\columnwidth}
        \centering
\includegraphics[width=1.15\textwidth]{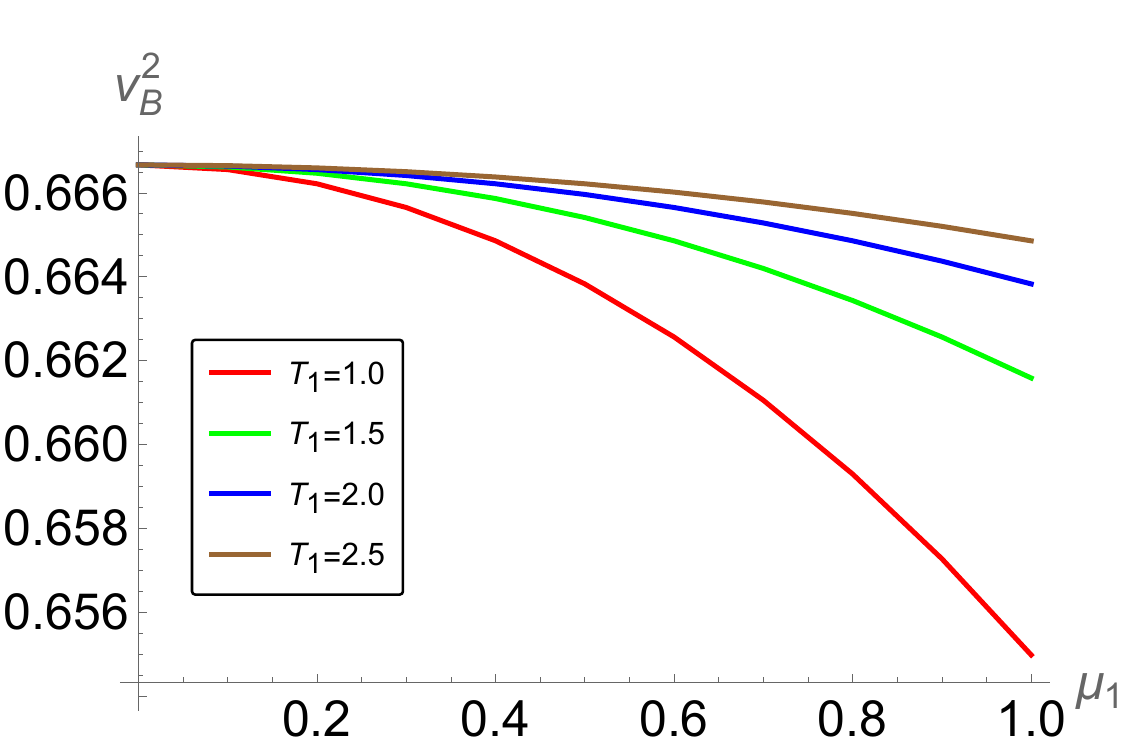}
        \caption*{(a) $v_B^2$ vs $\mu_1$ at various fixed values of the temperature $T_1$. }
    \end{subfigure}\hspace{10mm}
    \begin{subfigure}[b]{0.4\columnwidth}
        \centering
\includegraphics[width=1.15\textwidth]{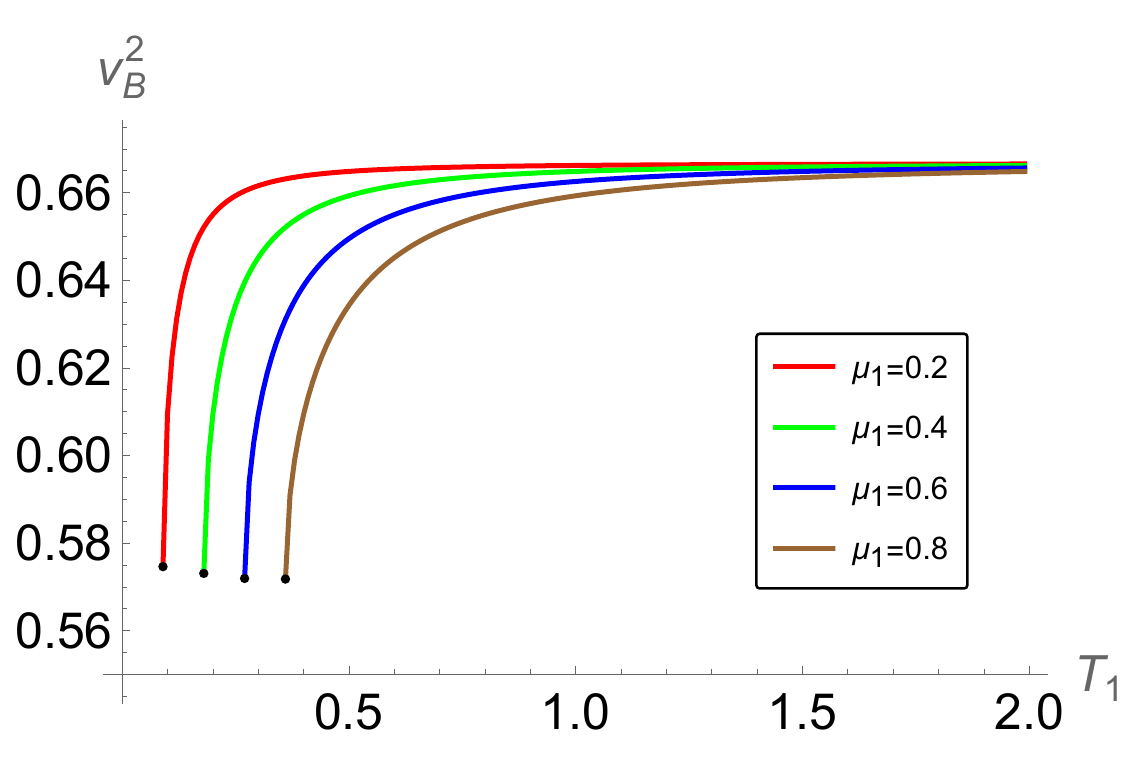}
        \caption*{(b) $v_B^2$ vs $T_1$ at various fixed values of the chemical potential $\mu_1$. The black dots denote the critical points.}
    \end{subfigure}\hspace{10mm}
    \caption{Variation of $v_B$ with respect to $\mu_1$ and $T_1$ for the 1RCBH model.}
    \label{fig:1RCBH}
\end{figure}

In Figure~\ref{fig:1RCBH}, we illustrate the chemical potential and temperature-dependent profiles of the butterfly velocity in the 1RCBH model. We find that $v_{B}^{2}$ monotonically decreases with $\mu_1$ for all temperatures. This suggests that the rate at which information about the perturbation propagates among the local degrees of freedom in the dual boundary theory decreases as one increases the chemical potential. In Figure~\ref{fig:1RCBH}, we have only plotted the thermodynamically stable solution. Note that the range of values of $\mu_1/T_1$ probed by the 1RCBH model is $[0, \pi/\sqrt{2}]$ and this is automatically taken care of, due to the presence of the square root ($\sqrt{\pi^2 T_{1}^{2}-2\mu_{1}^{2}}$) in \eqref{eqn:zhmuT1RCBH}, as is also evident from the figures. We find that, in the allowed range of parameters $\mu_1$ and $T_1$, $v_{B}^{2}$ always exhibits a monotonically decreasing profile with respect to $\mu_1$.

Similarly, $v_{B}^{2}$ is found to increase monotonically with $T_1$ for all chemical potential values. In particular, $v_{B}^{2}$ increases rapidly at low temperatures and then saturates to a constant value at high temperatures. This is an expected result, as one can expect that the local degrees of freedom would diffuse faster as one increases the temperature.  It is interesting to note that $v_B$ asymptotes to $v_B^S$, with $v_B^S$ being the butterfly velocity in the background of the AdS-Schwarzschild black brane, in the high-temperature limit ($T_1 \rightarrow \infty$) for all values of $\mu$, i.e., 
\begin{eqnarray}
    \label{eqnLvBSchw}
  v_{B}|_{T_1 \rightarrow \infty}  = v_B^S = \sqrt{\frac{d}{2(d-1)}} = \sqrt{\frac{2}{3}}\,.
\end{eqnarray}
Note that, in the case of chargeless plasma, by considering the AdS-Schwarzschild black brane, $v_{B}^{2}$ was found to be independent of temperature and takes a constant value \cite{Mezei:2016wfz}. With the chemical potential, its thermal structure becomes much richer, as is evident from Figure~\ref{fig:1RCBH}. For relativistic theories, $v_B^S$ gives the upper bound on the butterfly velocity \cite{Mezei:2016wfz}. It is interesting to see that this bound is always respected and only gets saturated in the high-temperature limit in the 1RCBH model. This might be because, at very high temperatures, all other scales get suppressed, leading to a scale symmetry. As we will see in the following sections, this result remains true in other holographic plasmas as well. 

\begin{figure}[h]
    \centering
    \begin{subfigure}[b]{0.4\columnwidth}
        \centering
\includegraphics[width=1.15\textwidth]{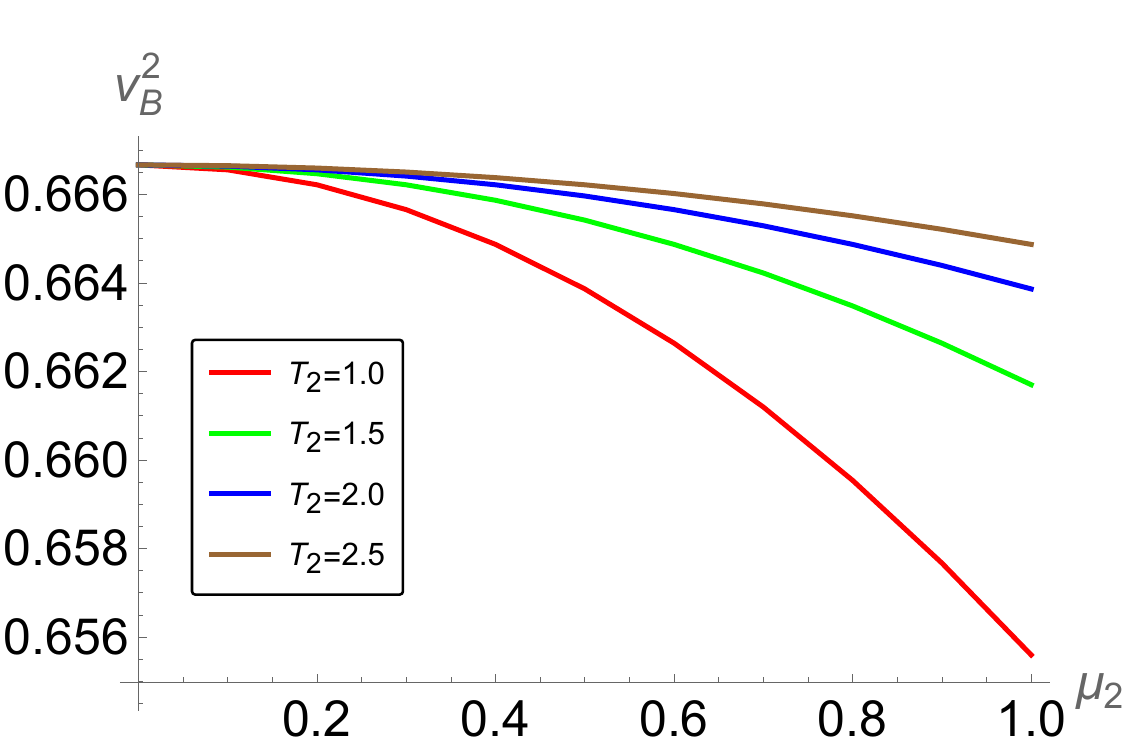}
        \caption*{(a) $v_B^2$ vs $\mu_2$ at various fixed values of the temperature $T_2$.}
    \end{subfigure}\hspace{10mm}
    \begin{subfigure}[b]{0.4\columnwidth}
        \centering
\includegraphics[width=1.15\textwidth]{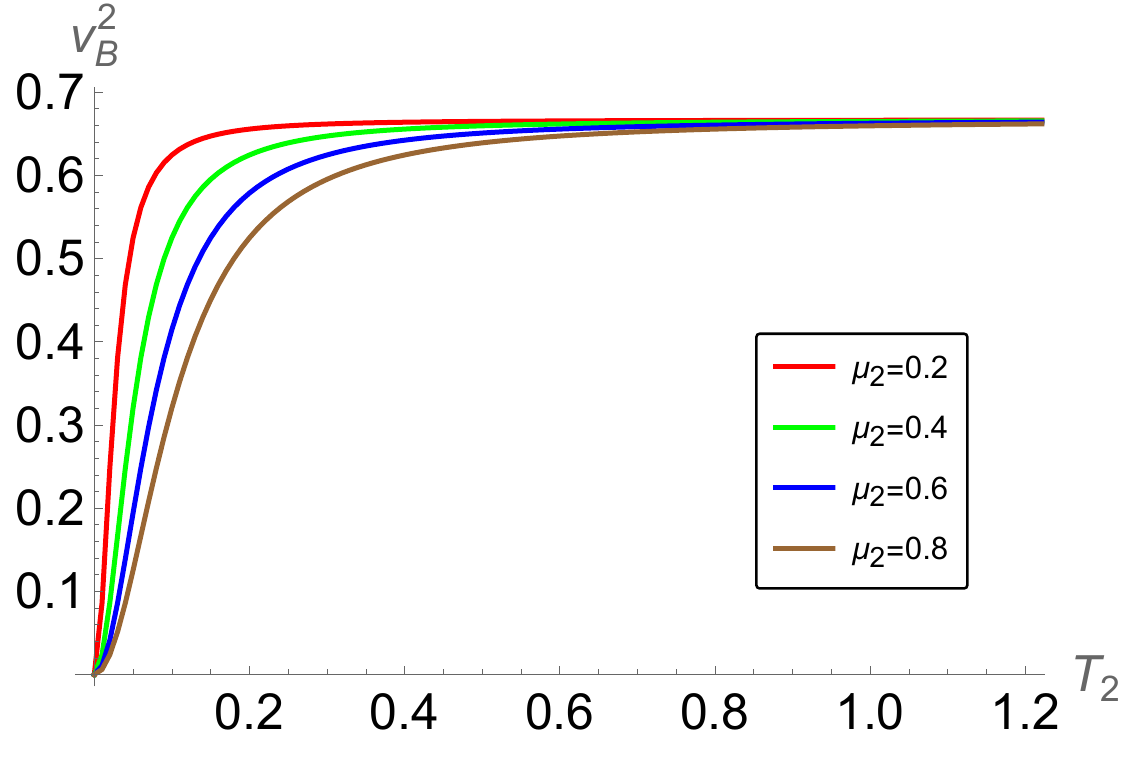}
        \caption*{(b) $v_B^2$ vs $T_2$ at various fixed values of the chemical potential $\mu_2$.}
    \end{subfigure}\hspace{10mm}
    \caption{Variation of $v_B$ with respect to $\mu_2$ and $T_2$ for the 2RCBH model.}
    \label{fig:2RCBH}
\end{figure}

The profile of the butterfly velocity in the 2RCBH model is shown in Figure~\ref{fig:2RCBH}. Most of the results of $v_{B}^{2}$ here are similar to the 1RCBH model. In particular, again $v_{B}^{2}$ decreases monotonically with the chemical potential at all fixed temperatures. Similarly, $v_{B}^{2}$ increases rapidly at low temperatures and then saturates to a constant value at high temperatures. Just like in the case of the 1RCBH model, $v_{B}$ for the 2RCBH model again asymptotes to $v_B^S$ in the high-temperature limit. However, there are also some differences. For instance, in contrast to the 1RCBH model, there are no restrictions on the allowed values of temperature and chemical potential in the 2RCBH model. Accordingly, one can probe $v_{B}^2$ in a much larger parameter space of $(T_2,~\mu_2)$ in 2RCBH model. This leads to some differences between the 1RCBH and 2RCBH models. In particular, at any fixed temperature, $v_{B}^2$ asymptotes to zero as the chemical potential is taken to infinity. This can be explicitly observed from Eq.~(\ref{eqn:vb2RCBHfinal}). However, this was not the case in the 1RCBH model, where $v_{B}^2$ attained a temperature-dependent finite minimum value. Similarly, at a fixed finite chemical potential, $v_{B}^2$ is bounded from below by a positive number in the 1RCBH model, whereas in the 2RCBH model, $v_{B}^2$ can be vanishingly small. Moreover, we further find that, for the same values of temperature and chemical potential, $v_{B}^2$ is higher in the 2RCBH model compared to the 1RCBH model, suggesting information propagates faster in the former compared to the latter model. 

\subsection{Calculation of $v_{B}$ using OTOC method}
The metric functions for the 1RCBH model are given in \eqref{eqn:metricfunctions1RCBH}. Substituting these metric functions into \eqref{eqn:relationofAandB} and using \eqref{eqn:zhminusuv}, we get the following expressions for $A(u,v)=A_1(u,v)$ and $B(u,v)=B_1(u,v)$ in the 1RCBH model:
\begin{eqnarray}
\label{eqn:A1RCBH}
  A_1(u,v) & = & \frac{\left[ 1 - \frac{z_h^{2}(z_h^{2} + Q_{1}^{2})}{(z_h-uv)^{2}((z_h-uv)^{2}+Q_{1}^{2})}\right]\left[(z_h-uv)^{2}\left(1+\frac{Q_{1}^{2}}{(z_h-uv)^{2}}\right)^{\frac{1}{3}}\right]}{2\alpha^{2}uv}\,, \nonumber \\
  B_1(u,v) & = & (z_h-uv)^{2}\left(1+\frac{Q_{1}^{2}}{(z_h-uv)^{2}}\right)^{\frac{1}{3}} \,.
\end{eqnarray}
Computing the limits of $A_1$ and $\partial_{u}\partial_{v}B_1$ as $u\rightarrow 0$, we get the following:
\begin{eqnarray}
    \label{eqn:AHorizon}
    A_1(0) & = & \lim_{u \to 0}  A_1(u,v) = -\frac{\left( Q_1^2 + 2 z_h^{2}\right)}{\left(1 + \frac{Q_1^2}{z_{h}^{2}}\right)^{\frac{2}{3}} \left(z_h \alpha^2 \right)}\,, \\
    \label{eqn:BHorizon}
    \partial_{u}\partial_{v}B_1(0) & = & \lim_{u \to 0} \partial_{u}\partial_{v}B_1(u,v) = -\frac{2 \left(2 Q_1^2 + 3z_h^{2}\right)}{3 z_h\left(1 + \frac{Q_1^2}{z_h^{2}}\right)^{\frac{2}{3}}}\,.   \end{eqnarray}
Substituting Eqs.~\eqref{eqn:AHorizon} and \eqref{eqn:BHorizon} into \eqref{eqn:chi}, we get
\begin{eqnarray}
\chi_1 = \sqrt{\frac{2 Q_1^2 + 3z_h^{2}}{Q_1^2 + 2z_h^{2}}} \alpha_1\,,
\end{eqnarray}
with $\alpha_1=2 \pi T_1$. Finally, putting everything into \eqref{eqn:VBOTOC}, we get the following expression of $v_{B}$:
\begin{equation}
    \label{eqn:vb1RCBHOTOC}
    v_{B}^{2} = \frac{2}{3}\left(\frac{z_h^{2}-\mu_{1}^{2}/2}{z_h^{2} -\mu_{1}^{2}/3}\right)\,,
\end{equation}
where we have traded $Q_{1}$ for $\mu_{1}$. The above expression matches with the analogous expression obtained for $v_{B}^{2}$ from the entanglement wedge method in the previous section [Eq.~\eqref{eqn:vb1RCBHEW}].

Similarly, we can compute $v_{B}$ for the 2RCBH model using the OTOC method. Now, $A(u,v)=A_2(u,v)$ and $B(u,v)=B_2(u,v)$ for the 2RCBH model are obtained as
\begin{eqnarray}
    \label{eqn:Afor2RCBH}
    A_2(u,v) & = & \frac{\left[1 - \frac{(z_h^{2} + Q_{2}^{2})^{2}}{((z_h-uv)^{2}+Q_{2}^{2})^{2}}\right]\left[(z_h-uv)^{2}\left(1+\frac{Q_{2}^{2}}{(z_h-uv)^{2}}\right)^{\frac{2}{3}}\right]}{2\alpha^{2}uv}\,, \\
    \label{eqn:Bfor2RCBH}
    B_2(u,v) & = & (z_h-uv)^{2}\left(1+\frac{Q_{2}^{2}}{(z_h-uv)^{2}}\right)^{\frac{2}{3}}\,.
\end{eqnarray}
The near-horizon limits are given as
\begin{eqnarray}
    \label{eqn:AHorizon2RCBH}
    A_2(0) & = & \lim_{u \to 0}  A_2(u,v) =  -\frac{2 z_h}{(1 + \frac{Q_2^2}{z_h^{2}})^{\frac{1}{3}} \alpha^2}\,, \\
    \label{eqn:BHorizon2RCBH}
    \partial_{u}\partial_{v}B_2(0) & = & \lim_{u \to 0} \partial_{u}\partial_{v}B_2(u,v) = -\frac{2 \left( Q_2^2 + 3 z_h^{2}\right)}{3 z_h\left(1 + \frac{Q_2^2}{z_h^{2}}\right)^{\frac{1}{3}}}\,.
\end{eqnarray}
Substituting the above equations into \eqref{eqn:chi}, we get
\begin{eqnarray}
\chi_2 = \frac{\sqrt{Q_2^2 + 3z_h^{2}}}{\sqrt{2}z_h} \alpha_2\,,
\end{eqnarray}
with $\alpha_2=2 \pi T_2$. Finally, from Eq.~\eqref{eqn:VBOTOC}, and expressing $Q_{2}$ in terms of $\mu_{2}$, we get the butterfly velocity
\begin{equation}
    \label{eqn:vb2RCBHOTOC}
    v_{B}^{2} = \frac{2}{3}\left(\frac{z_h^{2}}{z_h^{2} + \mu_{2}^{2}/6}\right)\,,
\end{equation}
which again takes the same form as in the case of the entanglement wedge method \eqref{eqn:vb2RCBHfinal}.

\subsection{Calculation of $v_{B}$ using pole-skipping method}
Now, we compute $v_B$ using the pole-skipping method. Here, we essentially show that Eq.~\eqref{eqn:RHSgeneralperturbEE} holds true for both 1RCBH and 2RCBH models.  The matter Lagrangian for both models is 
\begin{equation}
    \label{eqn:matterlagrangianEMD}    
    \mathcal{L}_{M}= \sqrt{-g}\left( -\frac{\tilde{f}(\phi)}{4} F_{\alpha \beta}F^{\alpha \beta} - \frac{1}{2} g^{\alpha \beta}(\partial_\alpha \phi)(\partial_\beta \phi) - V(\phi) \right)\,.
\end{equation}
The stress-energy tensor in terms of the matter Lagrangian is given as
\begin{equation}
    \label{eqn:TmunuinLm}
    T_{\alpha\beta}=2\left[\frac{1}{2}\mathcal{L}_{M}g_{\alpha\beta}-\frac{\partial\mathcal{L}_{M}}{\partial g^{\alpha\beta}}\right]\,.
\end{equation}
Explicitly, for the matter Lagrangian \eqref{eqn:matterlagrangianEMD}, the stress-energy tensor takes the form 
\begin{equation}
    \label{eqn:TmunuEMDLm}
    T_{\alpha\beta}=2\left[\frac{1}{2}\mathcal{L}_{M}g_{\alpha\beta} + \frac{1}{2}(\partial_\alpha \phi)(\partial_\beta \phi) + \frac{\tilde{f}(\phi)}{2} F_{\alpha\rho}F_{\beta\lambda}g^{\rho\lambda}\right]\,.
\end{equation}
From \eqref{eqn:TmunuEMDLm}, we get $T_{vz}$ as
\begin{equation}
    \label{eqn:TvzEMD}
    T_{vz}=2\left[\frac{1}{2}\mathcal{L}_{M}g_{vz} - \frac{\tilde{f}(\phi)}{2} (F_{vz})^2 g^{vz}\right]\,.
\end{equation}
Now, we differentiate \eqref{eqn:TmunuEMDLm} to obtain $\delta T_{\alpha\beta}$,
\begin{equation}
\label{eqn:DelTmunuEMDLm}
\begin{split}
\delta T_{\alpha\beta}=&\mathcal{L}_{M} \delta g_{\alpha\beta} + g_{\alpha\beta}\left(\frac{\delta \mathcal{L}_{M}}{\delta g^{\rho\lambda}}\delta g^{\rho\lambda} + \frac{\delta \mathcal{L}_{M}}{\delta \phi}\delta \phi+ \frac{\delta \mathcal{L}_{M}}{\delta A^{\rho}}\delta A^{\rho}\right)  + (\partial_\alpha \delta \phi)(\partial_\beta \phi) \\ & + (\partial_\alpha \phi)(\partial_\beta \delta \phi) + \tilde{f}'(\phi) F_{\alpha\rho}F_{\beta\lambda}g^{\rho\lambda}\delta \phi - \tilde{f}(\phi) F_{\alpha\mu}F_{\beta\sigma}g^{\mu\rho}g^{\lambda\sigma}\delta g_{\rho\lambda} \\ &+ \tilde{f}(\phi)g^{\rho\sigma} \left(F_{\alpha\rho}(\partial_\beta \delta A_{\sigma}-\partial_\sigma \delta A_{\beta}) +(\partial_\alpha \delta A_{\rho}-\partial_\rho \delta A_{\alpha})F_{\beta\sigma}\right)\,,
\end{split}
\end{equation}
from which $\delta T_{vv}$ at horizon turns out to be
\begin{equation}
\label{eqn:DelTvvEMDLm}
\begin{split}
\delta T_{vv}(z_{h})=\mathcal{L}_{M} \delta g_{vv}^{(0)} - \tilde{f}(\phi) (F_{vz})^2 (g^{vz}(z_{h}))^2 \delta g_{vv}^{(0)}\,.
\end{split}
\end{equation}
Here, we have used the fact that $g^{zz}(z_{h})=g_{vv}(z_{h})=0$. Substituting Eqs.~\eqref{eqn:TvzEMD} and \eqref{eqn:DelTvvEMDLm} into \eqref{eqn:RHSgeneralperturbEE} and using the fact that $g^{vz}(z_{h})=1/\sqrt{h(z_{h})f(z_{h})}$ and $g_{vz}(z_{h})= \sqrt{h(z_{h})f(z_{h})}$, it is straightforward to show that Eq.~\eqref{eqn:RHSgeneralperturbEE} holds true for both 1RCBH and 2RCBH models.

For completeness, we also compute the pole-skipping points of 1RCBH and 2RCBH models. Substituting metric functions from \eqref{eqn:metricfunctions1RCBH} into \eqref{eqn:TtermPSpointcomputated}, we get the following pole-skipping points in the 1RCBH model:
\begin{equation}
    \label{eqn:PSpoint1RCBH}
     \omega_{1*}=i\frac{Q_{1}^{2}+2z_{h}^{2}}{ \sqrt{Q_{1}^{2}+z_{h}^{2}}}=i2\pi T_1\,; ~~~ q_{1*} =i\sqrt{ \frac{(Q_{1}^{2}+2z_{h}^{2})(3z_{h}^{2}+2Q_{1}^{2})}{Q_{1}^{2}+z_{h}^{2}}}\,.
\end{equation}
From the above, we get a butterfly velocity that matches the results of the other two methods,
\begin{equation}
    \label{eqn:vb1RCBHPS}
    v_{B}^{2} = \frac{\omega_{1*}^{2}}{q_{1*}^{2}} =  \frac{2z_{h}^{2} + Q_{1}^{2}}{3z_{h}^{2} + 2Q_{1}^{2}}= \frac{2}{3}\left(\frac{z_{h}^{2}-\mu_{1}^{2}/2}{z_{h}^{2}-\mu_{1}^{2}/3}\right)\,.
\end{equation}
Similarly, the pole-skipping points of the 2RCBH model are given by
\begin{equation}
    \label{eqn:PSpoint2RCBH}
    \omega_{2*} =2iz_{h}=i2\pi T_2\,;~~~ q_{2*} = i\sqrt{ 2(3z_{h}^{2}+Q_{2}^{2})}\,.
\end{equation}
This leads to
\begin{equation}
    \label{VB2RCBHOTOC}
    v_{B}^{2} = \frac{2}{3}\left(\frac{z_{h}^{2}}{z_{h}^{2} + \mu_{2}^{2}/6}\right)\,.
\end{equation}
As expected, this again matches the expression obtained using the other two methods.

\section{Butterfly velocity in potential reconstruction based analytic bottom-up holographic QCD model}
\label{subsec:ourmodel}
\subsection{Background}
In this section, we turn our attention to a bottom-up holographic EMD model and investigate the butterfly velocity at finite temperatures and chemical potentials in the dual QCD deconfined plasma. As mentioned in the Introduction, bottom-up models generally mimic QCD properties more accurately as compared to their higher-dimensional string-theory-inspired top-down counterparts; therefore, it is interesting to analyze the butterfly velocity in bottom-up models as well. To make our analysis more complete, we will consider another sophisticated bottom-up EMD-based QCD model in the next section.

Here we consider the bottom-up holographic QCD model of \cite{Dudal:2017max, Bohra:2019ebj}. The most general version of this model is a solution to the EMD gravity (\ref{eqn:EMDaction})  containing Maxwell and dilaton matter fields. This model is based on the potential reconstruction technique \cite{Mahapatra:2018gig,Arefeva:2018hyo,Arefeva:2020byn,Mahapatra:2020wym,Priyadarshinee:2021rch,Priyadarshinee:2023cmi,Cai:2012xh,Daripa:2024ksg}, where the potential is self-consistently fixed from field equations. In particular, for the following ansatz for the metric, dilaton, and gauge fields, 
\begin{eqnarray}
    \label{eqn:ourmodelmetric}
   & &  ds^2 = e^{2A(z)}\left(-z^2G(z)dt^2+ \frac{dz^2}{z^2G(z)}+z^2(dx_1^2+dx^2+dx_3^2)\right)\,, \nonumber \\
  & & \phi = \phi(z)\,,~~~ \mathcal{A}_M = \Phi(z) \delta_{M}^0\,,  
\end{eqnarray}
one gets the following planar black hole solution:
\begin{align}
A(z) & = -\frac{a}{z^2}\,, ~~~~~~~~ \tilde{f}(z) = e^{-A(z)-\frac{c}{z^2}}\,, ~~~~~~~ \Phi(z) = \frac{\mu 
   \left(e^{\frac{c}{z_h^2}}-e^{\frac{c}{z^2}}\right)}{e^{\frac{c}{z_h^2}}-1} \,, \label{asol} \\
\phi(z) &= \frac{\sqrt{3} \sqrt{a \left(2 a+3 z^2\right)}}{z^2}
   - 3 \sqrt{\frac{3}{2}} \log \left(\frac{\sqrt{6} a^{3/2} z}{\sqrt{2} \sqrt{a^3 \left(2 a+3 z^2\right)}+2 a^2}\right) \,, \label{phisol} \\
G(z) &= 1-\frac{1+e^{\frac{3 a}{z^2}} \left(\frac{3 a}{z^2}-1\right)}{1+e^{\frac{3
   a}{z_h^2}} \left(\frac{3 a}{z_h^2}-1\right)} -\frac{c \mu ^2 \left(e^{\frac{3 a+c}{z_h^2}} \left(\frac{3
   a+c}{z_h^2}-1\right)+1\right)}{(3 a+c)^2 \left(e^{\frac{3 a}{z_h^2}}
   \left(\frac{3 a}{z_h^2}-1\right)+1\right)
   \left(e^{\frac{c}{z_h^2}}-1\right){}^2}    \\
   & + \frac{c \mu ^2 \left(e^{\frac{3 a+c}{z^2}} \left(\frac{3
   a+c}{z^2}-1\right)+1\right)}{(3 a+c)^2
   \left(e^{\frac{c}{z_h^2}}-1\right){}^2} -\frac{c \mu ^2 e^{\frac{3 a}{z^2}} \left(\frac{3 a}{z^2}-1\right)
   \left(e^{\frac{3 a+c}{z_h^2}} \left(\frac{3
   a+c}{z_h^2}-1\right)+1\right)}{(3 a+c)^2 \left(e^{\frac{3 a}{z_h^2}}
   \left(\frac{3 a}{z_h^2}-1\right)+1\right)
   \left(e^{\frac{c}{z_h^2}}-1\right){}^2}
   \,, \label{Hsol} \\
V(z) &= \frac{e^{-2 A(z)} }{2} \left(-z \left(\left(9 z A'(z)+11\right) g'(z)+z
   g''(z)\right)-6 g(z) \left(z^2 A''(z)+3 z^2 A'(z)^2+8 z
   A'(z)+4\right)\right) \,. \label{Vsol}
\end{align}
Here $\mu$ is the chemical potential, and $a$ and $c$ are model parameters that are fixed by taking inputs from real QCD. In particular, by demanding the deconfinement transition to be around $0.270~GeV$, as in the pure glue sector, fixes $a=0.15~\text{GeV}^2$, while the requirement of the linear Regge trajectory in the heavy-meson spectrum fixes $c=1.16~\text{GeV}^2$. Again, $z$ is the usual radial coordinate that runs from $z=z_h$ (horizon) to $z=\infty$ (boundary). The entropy and temperature of the black hole are given by
\begin{eqnarray}
\label{eqn:Tourmodel}
S &=& \frac{z_{h}^3~e^{-\frac{3a}{z_{h}^2}} V_3 }{4 G_5}\,, \nonumber \\
T &=& \frac{9 a^2
   e^{\frac{3 a}{z_h^2}}}{2 \pi  z_h \left(e^{\frac{3
   a}{z_h^2}} \left(3 a-z_h^2\right)+z_h^2\right)} + \frac{\mu ^2 e^{\frac{3 a}{z_h^2}} \left(\frac{9 a^2 c
   \left(e^{\frac{3 a+c}{z_h^2}} \left(3
   a+c-z_h^2\right)+z_h^2\right)}{(3 a+c)^2 \left(e^{\frac{3
   a}{z_h^2}} \left(3 a-z_h^2\right)+z_h^2\right)}-c
   e^{\frac{c}{z_h^2}}\right)}{2 \pi  z_h^3
   \left(e^{\frac{c}{z_h^2}}-1\right){}^2}\,.
\end{eqnarray}
Note that the above metric solution asymptotes to AdS at the boundary. This can be verified from the potential expression in (\ref{Vsol}). In particular, at the asymptotic boundary $z \rightarrow \infty$, $V(z)$ reduces to the value of
the cosmological constant expected in five-dimensional AdS spacetime. The dilaton potential is also bounded from above. Accordingly, this EMD model satisfies the Gubser criterion for a well-defined boundary theory \cite{Gubser:2000nd}. Moreover, the mass of the dilaton field respects the Breitenlohner-Freedman bound \cite{Breitenlohner:1982bm}, and the overall matter sector also satisfies the null energy condition \cite{Priyadarshinee:2021rch}. 

\begin{figure}[h]
    \centering
\includegraphics[width=0.5\textwidth]{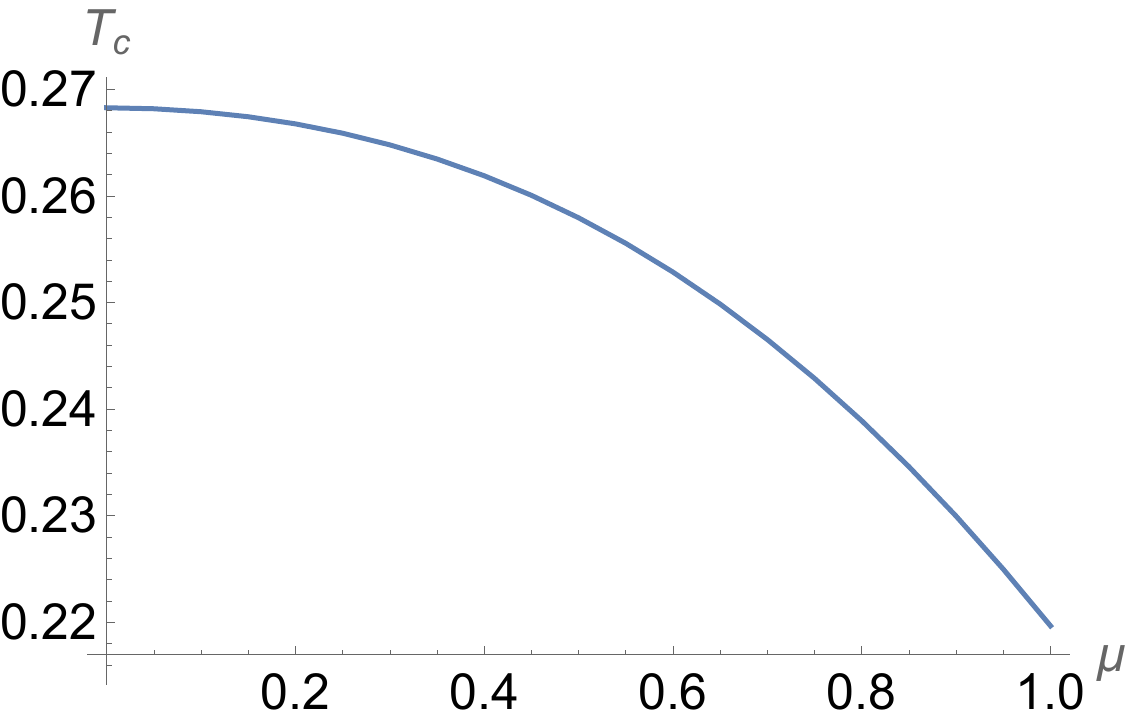}
    \caption{Variation of deconfinement transition temperature $T_c$ with respect to $\mu$. In units of MeV. }
    \label{fig:muvsTcmodel3}
\end{figure}

There is another solution to this gravity model. This second solution corresponds to thermal-AdS, which does not have a horizon. The thermodynamic phase structure of the above solution reveals that there are two black hole solutions (large and small) that exist only above a minimum temperature $T_{min}$; below $T_{min}$, these black hole solutions cease to exist. The large/small black hole solution has positive/negative specific heat and is thermodynamically stable/unstable. Interestingly, there exists a Hawking/Page type phase transition between thermal-AdS and large black hole solutions as the temperature is varied. In particular, the large black hole phase is thermodynamically favored at high temperatures, whereas the thermal-AdS phase is favored at low temperatures. The corresponding phase transition temperature $T_c$ is a chemical potential-dependent quantity, and in particular, it decreases as the chemical potential increases. This is shown in Figure~\ref{fig:muvsTcmodel3}. In the dual boundary theory, the thermal-AdS and large black hole phases correspond to the confined and deconfined phases. Accordingly, this EMD gravity model provides a self-consistent bottom-up holographic model for the deconfinement transition, which mimics real QCD behavior quite well. For these reasons, along with their semianalytic nature, potential reconstruction-based holographic models have been widely employed to investigate various properties of QCD. These applications include studies of energy loss and drag forces \cite{Arefeva:2018hyo}, transport properties of QGP \cite{Chen:2024epd, Jena:2025xcf}, heavy quarkonium melting and dissociation \cite{Jena:2024cqs, Jena:2022nzw}, chiral transition \cite{Bohra:2020qom}, etc. Since our main aim is to analyze the butterfly velocity and its thermal- and chemical-potential-dependent profile in the deconfined plasma phase, we will mainly concentrate on the large black hole solution in the next subsections.

It is also imperative to compare this holographic model with those of the 1RCBH and 2RCBH models considered earlier. From the QCD phenomenology point of view, the existence of the Hawking/Page phase transition allows us to consistently define the deconfinement transition temperature in the current model - a major improvement over the 1RCBH and 2RCBH models, where no such transition exists. Similarly, the dual confined phase, corresponding to the thermal-AdS background in the gravity side [obtained by taking the limit $z_h\rightarrow 0$ in Eq.~(\ref{Hsol})],  exhibits not only the area law of the Wilson loop but also the linear Regge trajectory for the heavy-meson mass spectrum. The latter feature, in particular, is absent in 1RCBH and 2RCBH models. These are well-established properties of the QCD confined phase, thereby making the current holographic model highly suitable from the QCD point of view. Having said that, it is also important to keep in mind that the potential reconstruction-based holographic QCD models should be understood as approximate models, as the on-shell dilaton potential depends mildly on the model parameters.

\subsection{$v_{B}$ from the entanglement wedge method}
To compute $v_{B}$ using the entanglement wedge method, we first need to recast Eq.~(\ref{eqn:ourmodelmetric}) in terms of Eq.~(\ref{eqn:generalmetric}). A straightforward comparison gives
\begin{equation}
\begin{split}  
\label{eqn:metricfunctionsourmodel}
    &g(z) = G(z)\,, \\ 
    &  h(z) = k(z) = z^{2} e^{-\frac{2a}{z^{2}}}\,, \\  
    & f(z) = \frac{e^{-\frac{2a}{z^{2}}}}{z^2}\,.
\end{split}      
\end{equation}
Computing the horizon values of the metric functions (and their derivatives) and substituting in \eqref{eqn:vbfinal}, we get the following analytic expression of the butterfly velocity in this bottom-up holographic QCD model: 
\begin{equation}
\label{eqn:vbourmodel}
v_B^2 = \frac{9 a^2
   e^{\frac{3 a}{z_h^2}}}{2 \pi  z_h \left(e^{\frac{3
   a}{z_h^2}} \left(3 a-z_h^2\right)+z_h^2\right)} + \frac{\mu ^2 e^{\frac{3 a}{z_h^2}} \left(\frac{9 a^2 c
   \left(e^{\frac{3 a+c}{z_h^2}} \left(3
   a+c-z_h^2\right)+z_h^2\right)}{(3 a+c)^2 \left(e^{\frac{3
   a}{z_h^2}} \left(3 a-z_h^2\right)+z_h^2\right)}-c
   e^{\frac{c}{z_h^2}}\right)}{2 \pi  z_h^3
   \left(e^{\frac{c}{z_h^2}}-1\right){}^2}\,.
\end{equation}
In the above equation, $v_B^2$ has been decomposed into two parts: one that explicitly depends on the chemical potential and the other, which is independent of it. The separated expressions clearly indicate that a nontrivial dependence of $v_B^2$  on $T$ and $\mu$ is expected in this model. Since it is not possible to analytically invert \eqref{eqn:Tourmodel} to express $z_h$ in terms of $\mu$ and $T$, we employ numerical methods to study $v_B^2$ as a function of these parameters. As mentioned above, this holographic model enjoys the Hawking/Page phase transition, with the large black hole phase being thermodynamically favored only at high temperature. Accordingly, below we mainly concentrate on the thermodynamically stable large black hole phase with temperature $T\geq T_c$, corresponding to a deconfined phase in the dual boundary theory. 

\begin{figure}[t]
    \centering
    \begin{subfigure}[b]{0.4\columnwidth}
        \centering
\includegraphics[width=1.15\textwidth]{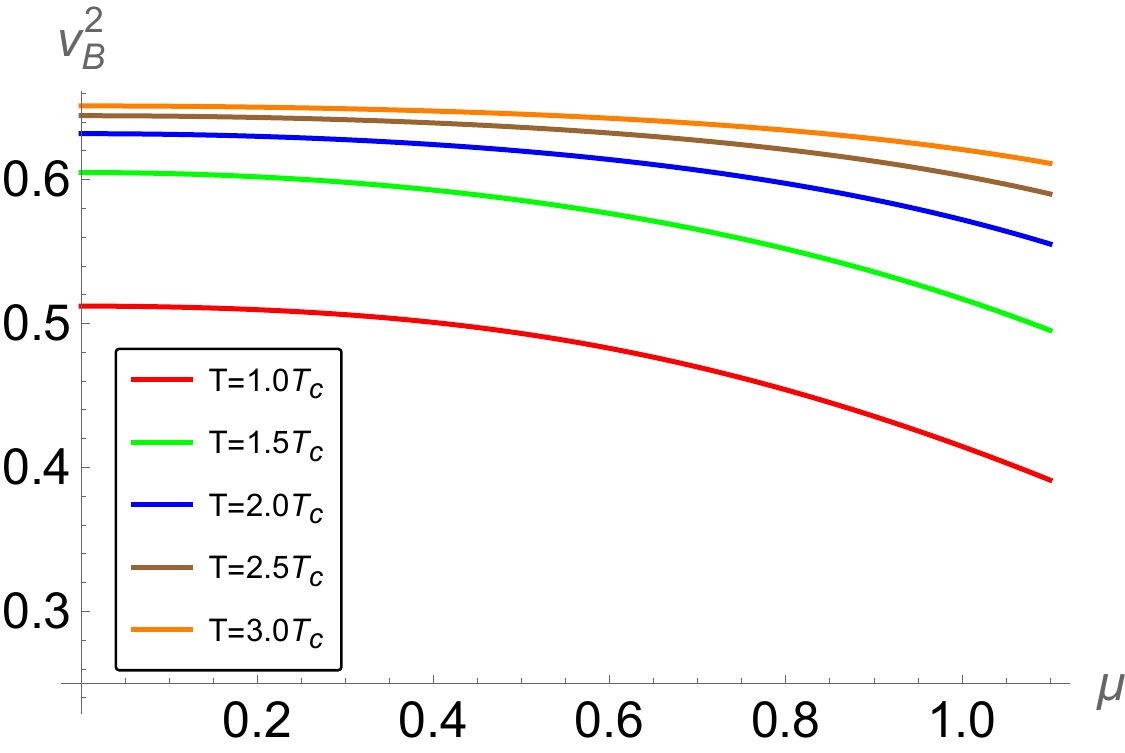} 
        \caption*{(a) $v_B^2$ vs $\mu$ at various fixed values of temperture $T/T_c$.}
    \end{subfigure}\hspace{10mm}
    \begin{subfigure}[b]{0.4\columnwidth}
        \centering
\includegraphics[width=1.15\textwidth]{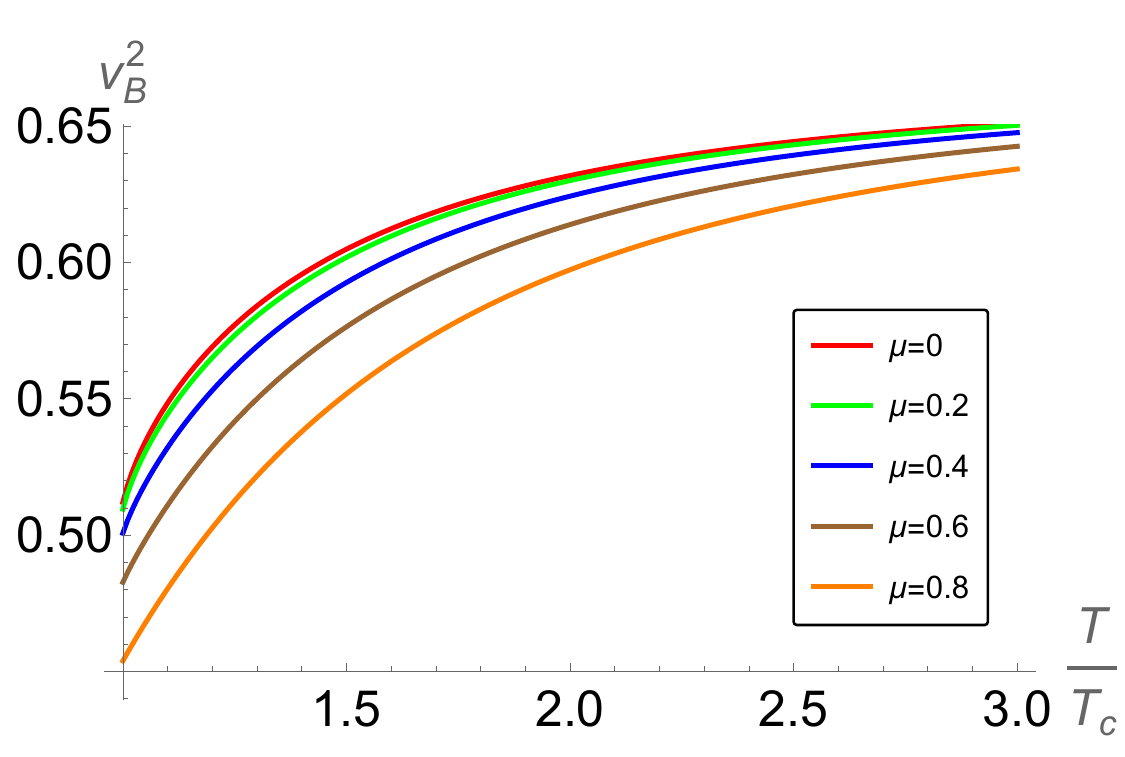}
        \caption*{(b) $v_B^2$ vs $T$ at various fixed values of the chemical potential.}
    \end{subfigure}\hspace{10mm}
    \caption{Variation of $v_B$ with respect to $\mu$ and $T$ for the analytic bottom-up holographic QCD model.}
    \label{fig:ourmodel}
\end{figure}

Our results for the butterfly velocity as a function of chemical potential and temperature are shown in Figure~\ref{fig:ourmodel}. Here we concentrate on the temperature range $T=T_c$ to $T=3.0~T_c$ and chemical potential range $\mu=0$ to $\mu=1.0$, which are relevant for the deconfined phase of QCD. We find that $v_B^2$ decreases monotonically with the chemical potential in the deconfined phase. This is true for all temperatures greater than the deconfined temperature. Similarly, for a fixed $\mu$, $v_B^2$ increases monotonically with temperature at low temperatures and then saturates to a constant value at high temperatures. Again, this constant value is $2/3$ and is independent of the chemical potential. This result can also be analytically observed from Eq.~(\ref{eqn:vbourmodel}). Note that in the large temperature limit $T\rightarrow \infty$, corresponding to $z_h\rightarrow \infty$, we get 
\begin{equation}
\label{eqn:vbourmodelasymp}
v_B^2 = \frac{2}{3} - \frac{6 a+\mu ^2}{9 z_h^2}+ \mathcal{O}(z_{h}^{-4})\,,
\end{equation}
suggesting that model parameters and chemical potential only give subleading corrections to $v_B^2$ in the large temperature limit. The above behavior of $v_B^2$ is quite similar to its behavior in the 1RCBH and 2RCBH models discussed earlier; i.e., the rate at which information about the perturbation propagates among the local degrees of freedom in the dual boundary deconfined phase decreases/increases with the chemical potential/temperature. Moreover, we have analyzed the butterfly velocity over a broad region of the temperature-chemical potential parameter space and consistently found that its behavior aligns with the pattern described above.

However, there are some important differences as well. In particular, note that for zero chemical potential, both 1RCBH and 2RCBH backgrounds reduce to the AdS-Schwarzschild background. Correspondingly, $v_B^2$ becomes constant (independent of temperature) at zero chemical potential in both 1RCBH and 2RCBH. In other words, since the scalar field in 1RCBH and 2RCBH models is sourced by the gauge field and therefore is a secondary hair in nature, it has no independent effects on the butterfly velocity. However, this is not the case in the current holographic model. Here, the scalar field is not sourced by the gauge field and is primary in nature. Accordingly, it also influences the properties of $v_B^2$ even when the chemical potential is zero. This can be explicitly observed from the red line in Figure~\ref{fig:ourmodel}, where one can see that $v_B^2$ varies nontrivially with the temperature and is not a constant.

\subsection{$v_{B}$ from the OTOC method}
Substituting the metric functions [given by \eqref{eqn:metricfunctionsourmodel}] and \eqref{eqn:zhminusuv} into \eqref{eqn:relationofAandB}, we get the following complicated expressions of $A(u,v)$ and $B(u,v)$:
\begin{equation}   
 \begin{aligned}
A(u,v) &= \frac{\mu ^2 \left(z_h-uv\right){}^2 \left(\frac{c \left(e^{\frac{3 a+c}{\left(z_h-uv\right){}^2}} \left(\frac{3
   a+c}{\left(z_h-uv\right){}^2}-1\right)+1\right)}{(3 a+c)^2 \left(1-e^{\frac{c}{z_h^2}}\right){}^2}-\frac{c \left(e^{\frac{3 a+c}{z_h^2}}
   \left(\frac{3 a+c}{z_h^2}-1\right)+1\right) e^{-\frac{2 a}{\left(z_h-uv\right){}^2}} \left(e^{\frac{3
   a}{\left(z_h-uv\right){}^2}} \left(\frac{3 a}{\left(z_h-uv\right){}^2}-1\right)+1\right)}{(3 a+c)^2 \left(e^{\frac{3 a}{z_h^2}}
   \left(\frac{3 a}{z_h^2}-1\right)+1\right) \left(1-e^{\frac{c}{z_h^2}}\right){}^2}\right)}{2 \alpha ^2
   uv}+ \\
   &\frac{\left(z_h-uv\right){}^2 \left(\frac{1}{(3 a+c)^2 \left(1-e^{\frac{c}{z_h^2}}\right){}^2}-\frac{e^{-\frac{2
   a}{\left(z_h-uv\right){}^2}} \left(e^{\frac{3 a}{\left(z_h-uv\right){}^2}} \left(\frac{3
   a}{\left(z_h-uv\right){}^2}-1\right)+1\right)}{e^{\frac{3 a}{z_h^2}} \left(\frac{3 a}{z_h^2}-1\right)+1}\right)}{2 \alpha ^2 uv}\,,
 \end{aligned}
\end{equation}
\begin{eqnarray}
    \label{eqn:Bforourmodel}
& B(u,v) = e^{-\frac{2a}{(z_h-uv)^{2}}}(z_h-uv)^{2}\,.
\end{eqnarray}
Computing the near-horizon limits, we get
\begin{eqnarray}
\small
\label{eqn:AHorizonourmodel} 
&  & A(0) = \frac{2 e^{\frac{a}{z_h^2}} v \left(c e^{\frac{c}{z_h^2}} \mu^2 - \frac{9 a^2 \left((3 a + c)^2 \left(-1 + e^{\frac{c}{z_h^2}}\right)^2 z_h^2 + c \left(z_h^2 + e^{\frac{3 a + c}{z_h^2}} (3 a + c - z_h^2)\right) \mu^2\right)}{(3 a + c)^2 \left(z_h^2 + e^{\frac{3 a}{z_h^2}} (3 a - z_h^2)\right)}\right)}{\left(-1 + e^{\frac{c}{z_h^2}}\right)^2 z_h^3}\,, \\
& & \partial_{u}\partial_{v}B (0) = -\frac{2 e^{-\frac{2a}{z_h^2}} \left(2a + z_h^2\right)}{z_h}\,.
\end{eqnarray}
Substituting the above equations into \eqref{eqn:VBOTOC} and simplifying, we get the same expression for the butterfly velocity as \eqref{eqn:vbourmodel}. The results from the two methods again match.

\subsection{$v_{B}$ from the pole-skipping method}
The matter Lagrangian of the holographic model considered in this section is the same as \eqref{eqn:matterlagrangianEMD} of the 1RCBH and 2RCBH models considered in the previous section, as both models are constructed from the same EMD action. This implies that the Eq.~\eqref{eqn:RHSgeneralperturbEE} also holds true for this model. Accordingly, the pole-skipping points are again given by Eq.~(\ref{eqn:TtermPSpointcomputated}). Substituting the metric functions from \eqref{eqn:metricfunctionsourmodel} into \eqref{eqn:TtermPSpointcomputated}, we obtain the following expression of the pole-skipping points in the current holographic model:
\begin{eqnarray}
\label{eqn:PSpointourmodelomega}
 \omega_{*}^{2} & = & -\left(\frac{9 a^2
   e^{\frac{3 a}{z_h^2}}}{  z_h \left(e^{\frac{3
   a}{z_h^2}} \left(3 a-z_h^2\right)+z_h^2\right)} + \frac{\mu ^2 e^{\frac{3 a}{z_h^2}} \left(\frac{9 a^2 c
   \left(e^{\frac{3 a+c}{z_h^2}} \left(3
   a+c-z_h^2\right)+z_h^2\right)}{(3 a+c)^2 \left(e^{\frac{3
   a}{z_h^2}} \left(3 a-z_h^2\right)+z_h^2\right)}-c
   e^{\frac{c}{z_h^2}}\right)}{ z_h^3
   \left(e^{\frac{c}{z_h^2}}-1\right){}^2}\right)^{2}\,,
\end{eqnarray}
\begin{eqnarray}
 q_{*}^{2} & = & \frac{-54 a^2 e^{\frac{3 a}{z_h^2}} \left(2 a+z_h^2\right)}{z_h^2 \left(e^{\frac{3 a}{z_h^2}} \left(3
   a-z_h^2\right)+z_h^2\right)}-\frac{6 \mu ^2 e^{\frac{3 a}{z_h^2}} \left(2 a+z_h^2\right) 
    \left(\frac{9 a^2 c \left(e^{\frac{3 a+c}{z_h^2}} \left(3a+c-z_h^2\right)+z_h^2\right)}{(3 a+c)^2 \left(e^{\frac{3 a}{z_h^2}} \left(3 a-z_h^2\right)+z_h^2\right)}-c e^{\frac{c}{z_h^2}}\right)}{z_h^4
   \left(e^{\frac{c}{z_h^2}}-1\right){}^2}.
   \label{eqn:PSpointourmodelq}
\end{eqnarray}
Substituting the above equations into \eqref{eqn:generalPSlambdaandvb} and simplifying, we recover the butterfly velocity expression previously obtained in \eqref{eqn:vbourmodel}, confirming the consistency of the result.
\begin{figure}[h]
    \centering
    \begin{subfigure}[b]{0.47\textwidth}
        \centering
        \includegraphics[width=\textwidth]{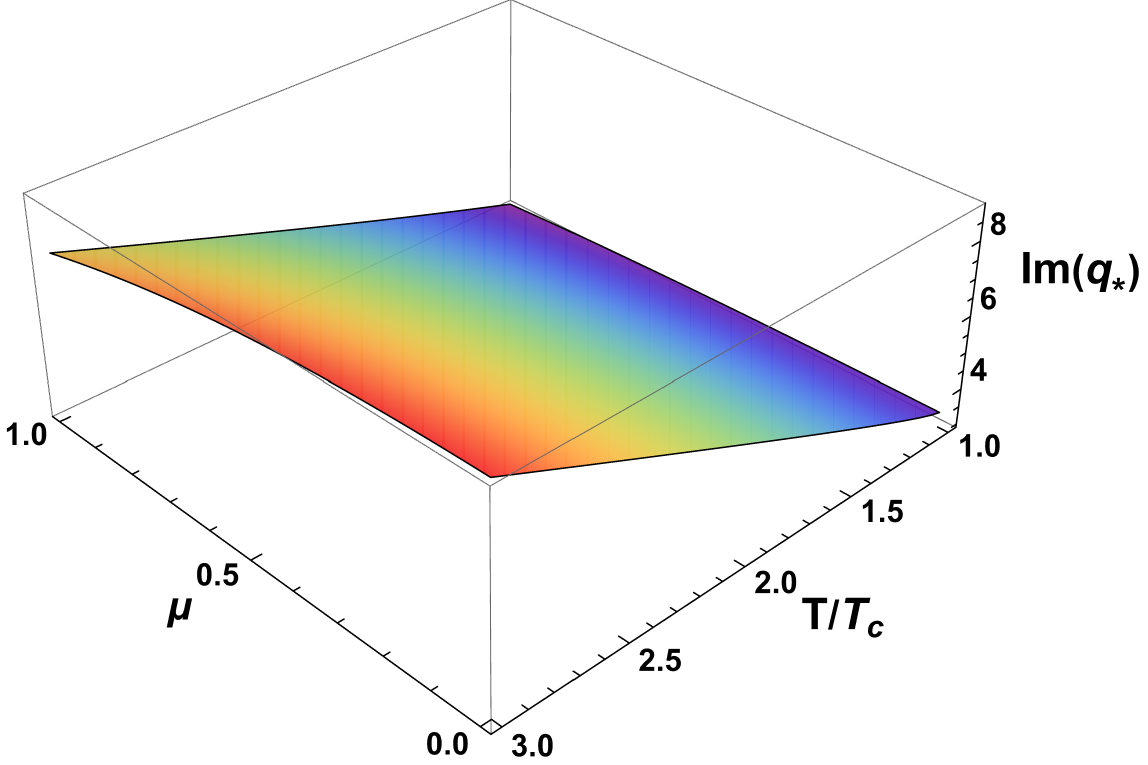}
        \caption{Im($q_{*}$) as a function of $\mu$ and $T/T_c$.}
        \label{fig:Imq}
    \end{subfigure}
    \hfill
    \begin{subfigure}[b]{0.47\textwidth}
        \centering
        \includegraphics[width=\textwidth]{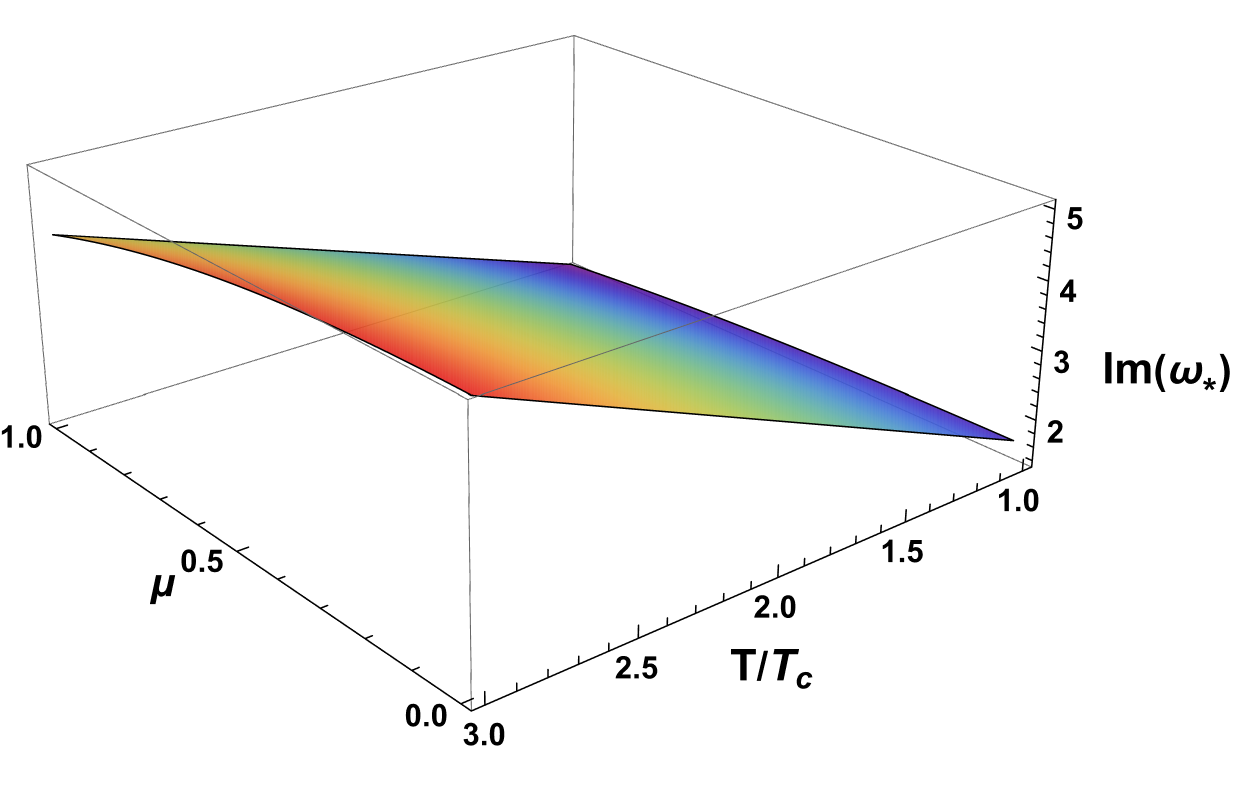} 
        \caption{Im($\omega_{*}$) as a function of $\mu$ and $T/T_c$. }
        \label{fig:Imw}
    \end{subfigure}
    \caption{Pole-skipping points $q_*$ and $\omega_*$ as a function of temperature and chemical potential.}
    \label{fig:combined}
\end{figure}

For completeness, we also analyze the overall structure of the pole-skipping points. This is shown in Figure \ref{fig:combined}, where the imaginary parts of $q_{*}$ and $\omega_{*}$ as a function of $\mu$ and $T/T_c$ are illustrated. Again, these special points are smooth and monotonic in the parameter space of temperature and chemical potential. Both points increase linearly with $T$. Also, they exhibit marginal dependence on $\mu$, which is more pronounced near the deconfinement temperature. 

\section{Numerical Bottom-up Holographic QCD model}
\label{subsec:Bottomu2p}
\subsection{Background}
In this section, we turn our attention to another bottom-up holographic QCD model, particularly to the model of \cite{Rougemont:2015wca}, to probe the temperature- and chemical-potential-dependent profile of the butterfly velocity. Again, this model is based on the EMD gravity (\ref{eqn:EMDaction}). Unlike the EMD models considered in the previous sections, this model is completely numerical and requires numerical integration of the gravity equations of motion from the horizon to the boundary to obtain boundary field theory information. Since the foundational work of \cite{DeWolfe:2010he}, numerous numerical EMD-based AdS/QCD models have been developed, including \cite{Rougemont:2015wca}, and others, such as \cite{DeWolfe:2011ts,Critelli:2017oub,Finazzo:2016mhm,Knaute:2017lll,Cai:2022omk, Jokela:2024xgz}. These holographic models share a common strategy: the dilaton potential  $V(\phi)$ and the gauge-kinetic function $\tilde{f}(\phi)$ are chosen to match lattice QCD results for observables like entropy density, pressure, and susceptibility at zero chemical potential. Once the forms of $V(\phi)$ and $\tilde{f}(\phi)$ are calibrated at zero chemical potential, results with nonzero chemical potential constitute genuine holographic predictions. Over the past decade, such numerical models have become key tools for exploring QCD via holography. In what follows, we primarily examine the model of \cite{Rougemont:2015wca}, though our analysis of the butterfly velocity can be readily extended to other similar frameworks.
 
Before we explicitly analyze the butterfly velocity in the model of \cite{Rougemont:2015wca} and other variants of it, let us briefly differentiate it from the models considered previously. Apart from the obvious difference at the solution level, which is now completely numerical as opposed to the previous analytical models, this model is mainly suitable for studying finite-temperature QCD features in the deconfined phase. In particular, the gravity equations are solved to construct numerical black holes in the bulk. However, setting up the numerical procedure to compute confined properties requires implementing the initial value problem deep in the bulk (strictly speaking, at $z=0$), which poses serious numerical challenges. Consequently, one faces problems in studying important observables, such as the Wilson and Polyakov loops, in the confined phase here.\footnote{It is worth noting that, by choosing an appropriate dilaton potential, one can construct a four-dimensional gauge theory with a nontrivial renormalization group (RG) flow at zero temperature, which runs from an ultraviolet (UV) fixed point to an infrared (IR) fixed point. In the dual gravity description, this RG flow corresponds to a domain-wall geometry that interpolates between two AdS spaces \cite{Attems:2016ugt}.} This is in contrast to the previous models, where these observables can be computed explicitly. Similarly, unlike the holographic model of Section~\ref{subsec:ourmodel}, it is difficult to directly obtain the deconfinement temperature here, as the gravity model does not exhibit the Hawking/Page phase transition. Accordingly, one can not consistently define the deconfinement temperature and rely on indirect techniques for the same.

Despite these limitations, such numerical models have become central to phenomenological studies of QCD via holography. They have been employed to determine the QCD critical point \cite{Critelli:2017oub}, study anisotropic heavy-quark drag forces and Langevin diffusion coefficients \cite{Finazzo:2016mhm}, analyze the entanglement and complexity structure of QCD phases \cite{Knaute:2017lll,Zhang:2017nth}, and explore the behavior of QCD in the presence of a magnetic field \cite{Rougemont:2015oea}, among other applications. More recently, these models have also been used to construct the QCD equation of state via neural networks \cite{Cai:2024eqa}, and to investigate stochastic gravitational waves and primordial black hole (PBH) production in the early Universe \cite{He:2022amv}. Here, we use such models to analyze the chaotic aspect of QCD. 

Now we briefly discuss its underlying mathematical structure and numerical implementation. In this bottom-up EMD model, the following form of the dilaton potential and coupling function is considered:
\begin{eqnarray}
  & & V(\phi) = -12 \cosh{(c_1 \phi)} + c_2\phi^2 + c_{4}\phi^4 + c_{6}\phi^6\,, \nonumber\\
 & & \tilde{f}(\phi) = \frac{\sech{(b_1 \phi - b_2)}}{3\sech{(b_2)}} + \frac{2}{3}e^{-b_3 \phi}\,,
 \label{potentialmodel4}
\end{eqnarray}
with $c_1=0.606$, $c_2=0.703$, $c_4=-0.1$, $c_6=0.0034$, $b_1=1.2$, $b_2=0.69$, and $b_3=100$. The free parameters $c_i$ and $b_i$  are again chosen to match with physical properties of realistic QCD.  The ansatz
\begin{eqnarray}
\label{metricmodel4}
  & & ds^2=e^{2 \eta(z)}\left[-\mathcal{G}(z)dt^2 + dx_{1}^{2}+dx_{2}^{2}+dx_{3}^{2} \right] +\frac{e^{2\tau(z)}dz^2}{\mathcal{G}(z)}\,, \nonumber \\
  & & \phi=\phi(z)\,,~~~\mathcal{A}_M dx^M = \Phi(z)dt\,,
\end{eqnarray}
represents an asymptotically $AdS_5$ spacetime with
boundary at $z\rightarrow \infty$ and defines a black hole horizon by $\mathcal{G}(z_h)=0$. Following \cite{DeWolfe:2010he}, here we mainly work in the gauge $\tau(z)=0$. 
 With ansatz~(\ref{metricmodel4}) for the metric, dilaton, and gauge fields, one gets the following four independent equations of motion:
\begin{eqnarray}
\label{eqn:EOMbottomup}
& & \eta '' + \frac{1}{6} \phi'^2 = 0\,, \\
& & \mathcal{G}'' + 4 \eta' \mathcal{G}' - e^{-2\eta} \tilde{f}(\phi)\Phi'^2 = 0\,, \\
& & \Phi'' + 2 \eta' \Phi' + \frac{d \log \tilde{f}(\phi)}{d \phi} \phi' \Phi' = 0\,, \\
& & \phi'' + \left(4 \eta' + \frac{\mathcal{G'}}{\mathcal{G}} \right)\phi' -\frac{1}{\mathcal{G}} \frac{\partial V_{eff}}{\partial \phi} = 0\,,
\end{eqnarray}
and a zero energy constraint equation
\begin{equation}
    \label{eqn:zeroenergyconstraint}
    \mathcal{G}(24 \eta'^2 - \phi'^2) + 6\eta'\mathcal{G}' + 2 V(\phi) + e^{-2 \eta} f(\phi) \Phi'^2  = 0\,. 
\end{equation}
Here,
\begin{equation}
    V_{\text{eff}}(\phi,r) \equiv V(\phi) - \frac{1}{2} e^{-2 \eta} \tilde{f}(\phi) \Phi'^2\,,
\end{equation}
and the prime denotes derivative with respect to the radial coordinate $z$. The complicated forms of the $V(\phi)$ and $\tilde{f}(\phi)$ essentially make the gravity system analytically intractable, and one therefore has to resort to numerics. Assuming that all field variables depend only on the radial coordinate $z$ and are smooth at the horizon $z=z_h$, they can be Taylor expanded near the horizon as 
\begin{eqnarray}
  & & \mathcal{Y}=\sum_{n=0}^{\infty}\mathcal{Y}_n (z-z_h)^n\,,
\end{eqnarray}
where $\mathcal{Y}=\{\phi, \Phi, \mathcal{G}, \eta \}$. By rescaling the radial $z$ and time $t$ coordinates, one can fix the location of the horizon $z_h=0$ and $\mathcal{G}_1=1$. Similarly, $\mathcal{G}_0=0$ by definition and $\Phi_0=0$ is required to have a well-defined gauge field at the horizon. Also, $\eta_0=0$
can be fixed by rescaling $(t, \vec{x})$ by a common factor. It is straightforward to see that given a two-parameter initial condition, $\{\phi_0, \Phi_1\}$, all other remaining coefficients in the near-horizon expansions of the bulk fields can be systematically determined by recursively solving the equations of motion order by order. Note that $\phi_0$ is the value of the scalar field at the horizon and $\Phi_1$ is essentially the electric field at the horizon. Given the coefficients $(\Phi_h, \phi_h)$, one can set up the boundary conditions at the horizon and then numerically integrate the equations of motion from the horizon to the boundary. 

On the other hand, the UV expansion of bulk fields at the asymptotic boundary $z\rightarrow \infty$ is \cite{DeWolfe:2010he}
\begin{eqnarray}
& & \phi(z) = \phi_A e^{-\nu \alpha(z)} ( 1+ a_{2\nu} e^{-2 \nu \alpha(z)} +\dots) + \phi_B e^{-\Delta_\phi \alpha(z)}+\dots \,, \nonumber \\
& & \Phi = \Phi_{0}^{\text{far}} + \Phi_{2}^{\text{far}} e^{-2 \alpha(z)} + \dots \,, \nonumber \\
& & \mathcal{G}(z) = \mathcal{G}_{0}^{\text{far}} + \mathcal{G}_{4}^{\text{far}} e^{-4 \alpha(z)} + \dots \,,  \nonumber \\
& & \eta(z) = \alpha(z) + \eta_{2\nu}^{\text{far}} e^{-2 \nu \alpha(z)}+\dots \,,
\label{nearboundexpmodel4numerical}
\end{eqnarray}
where $\alpha(z)=z/\sqrt{\mathcal{G}_{0}^{\text{far}}}+\eta_{0}^{\text{far}}$ and $\nu=4-\Delta_\phi$, with $\Delta_\phi \approx 3$, being the dimension of the operator dual to $\phi$. 

The coordinates used in Eq.~(\ref{metricmodel4}) are useful to numerically solve the EMD equations of motion. To make a meaningful comparison of solutions and calculate physical observables of dual theory, one should ideally arrange for the source terms $\phi_A$  to take the same value. This is generally done through a coordinate transformation \cite{DeWolfe:2010he}. In particular, suppose we numerically obtain a solution characterized by a positive value of $\phi_A$. We then aim to apply a coordinate transformation to this solution to express it in the following form:
\begin{eqnarray}
 & &   d \tilde{s}^2 = e^{2 \tilde{\eta}(\tilde{r})} (-\tilde{\mathcal{G}}(\tilde{r})d \tilde{t}^2 + d \tilde{\vec{x}}^2) + \frac{d \tilde{r}^2}{\tilde{\mathcal{G}}(\tilde{r})}\,, \nonumber \\ 
& & \tilde{\phi}=\tilde{\phi}(\tilde{z})\,,~~~\tilde{\mathcal{A}}_M d\tilde{x}^M = \tilde{\Phi}(\tilde{z})d\tilde{t}\,,
\label{metricmodel4tilde}
\end{eqnarray}
having near-boundary expansion
\begin{eqnarray}
    & & \tilde{\eta} (\tilde{z}) = \tilde{z} + \mathcal{O} (e^{-2 \nu \tilde{z}})\,, \nonumber \\ 
    & & \tilde{\mathcal{G}} (\tilde{z}) = 1 + \tilde{\mathcal{G}}_{4}^{\text{far}} e^{- 4 \tilde{z}} + \mathcal{O} (e^{-(4 + 2 \nu)\tilde{z}})\,, \nonumber \\ 
    & & \tilde{\Phi}(\tilde{z}) = \tilde{\Phi}_{o}^{\text{far}} + \tilde{\Phi}_{2}^{\text{far}} e^{-2 \tilde{z}} + \mathcal{O} (e^{-(2 + \nu)\tilde{z}})\,, \nonumber \\ 
    & & \tilde{\phi} (\tilde{z}) = e^{-\nu \tilde{z}} + \mathcal{O} (e^{-2 \nu \tilde{z}})\,.
    \label{eqn:tildefunctions1}
\end{eqnarray}
To distinguish these coordinates, following \cite{Rougemont:2015wca}, we refer to the coordinates with a tilde as standard coordinates and those without a tilde as numerical coordinates. Note that the blackening function $\tilde{\mathcal{G}}(\tilde{z})$ goes to $1$ at the asymptotic boundary in the standard coordinates, thereby allowing us to compute physical quantities, such as the temperature or chemical potential, using the standard holographic techniques. Now, setting $ds^2=d\tilde{s}^2$, $\tilde{\phi}(\tilde{z})=\phi(z)$, $\tilde{\Phi}(\tilde{z})d\tilde{t}=\Phi(z)dt$, and comparing the near-boundary expansion [Eqs.~(\ref{nearboundexpmodel4numerical}) and (\ref{eqn:tildefunctions1})], one gets the following relations between the standard and numerical coordinates:
\begin{eqnarray}
\tilde{t} = \phi_A^{1/\nu} \sqrt{\mathcal{G}_0^{\text{far}}}\, t\,,~~~~\tilde{\vec{x}} = \phi_A^{1/\nu} \vec{x}\,,~~~~
\tilde{z} = \alpha(z) - \log\left( \phi_A^{1/\nu} \right)\,.
\end{eqnarray}
Similarly, we have
\begin{eqnarray}
\tilde{\eta}(\tilde{z}) = \eta(z) - \log(\phi_A^{1/\nu})\,,~~~~
\tilde{\mathcal{G}}(\tilde{z}) = \frac{1}{\mathcal{G}_0^{\text{far}}} \mathcal{G}(z)\,,~~~~
\tilde{\Phi}(\tilde{z}) = \frac{1}{\phi_A^{1/\nu} \sqrt{\mathcal{G}_0^{\text{far}}}} \, \Phi(z)\,.
\end{eqnarray}
The above equations further imply that
\begin{eqnarray}
\tilde{\Phi}_0^{\text{far}} = \frac{\Phi_0^{\text{far}}}{\phi_A^{1/\nu} \sqrt{\mathcal{G}_0^{\text{far}}}}\,,~~~~
\tilde{\Phi}_2^{\text{far}} = \frac{\Phi_2^{\text{far}}}{\phi_A^{3/\nu} \sqrt{\mathcal{G}_0^{\text{far}}}}\,,~~~~
\tilde{\mathcal{G}}_4^{\text{far}} = \frac{\mathcal{G}_4^{\text{far}}}{\phi_A^{4/\nu} \mathcal{G}_0^{\text{far}}}.
\end{eqnarray}
Using the above relations, one can recast the thermodynamic quantities of the dual boundary theory, such as the temperature $T$, chemical potential $\mu$, entropy density $S$, charge density $\rho$, etc., in terms of the coefficients of the near-boundary expansions in the numerical coordinates
\begin{eqnarray}
T &=& \frac{e^{\tilde{\eta}(\tilde{z}_h)}}{4\pi} \left( \frac{d\tilde{\mathcal{G}}}{d\tilde{z}} \right)_{\tilde{z} = \tilde{z}_h}
= \frac{1}{4\pi } \frac{1}{\phi_A^{1/\nu} \sqrt{\mathcal{G}_0^{\text{far}}}} \Lambda\,, \\ \nonumber
\mu &=& \tilde{\Phi}_0^{\text{far}} = \frac{\Phi_0^{\text{far}}}{ \phi_A^{1/\nu} \sqrt{\mathcal{G}_0^{\text{far}}}}\Lambda\,, \\ \nonumber
S & = & \frac{2\pi}{\kappa^2}e^{3\tilde{\eta}(\tilde{z}_h)} = \frac{2\pi}{\kappa^2} \frac{1}{\phi_A^{3/\nu}}\Lambda^3, \\ \nonumber
\rho & = & - \frac{\tilde{\Phi}_2^{\text{far}}}{\kappa^2} = - \frac{ \Phi_{2}^{\text{far}}}{ \kappa_{5}^{2} \phi_{A}^{3/\nu} \sqrt{\mathcal{G}_{0}^{\text{far}}}} \Lambda^3\,.
\end{eqnarray}
Here the energy scale $\Lambda \approx 831$~MeV  is introduced to convert quantities computed from black hole physics - originally expressed in units of the asymptotic AdS radius - into physical units of the boundary field theory. Therefore, by taking different profiles of the near-horizon data $(\phi_0, \Phi_1)$, we can compute the boundary data, such as $\{\phi_A, \mathcal{G}_{0}^{\text{far}}\}$, numerically and compute the boundary field theory observables. For more technical details of this numerical procedure, see \cite{Rougemont:2015wca, DeWolfe:2010he}. 

\subsection{$v_{B}$ from the entanglement wedge method}
Comparing the standard coordinate metric \eqref{metricmodel4tilde} with the general metric \eqref{eqn:generalmetric}, we have
\begin{equation}
\begin{split}  
\label{metriccompmodel4}
    &g(z) = \tilde{\mathcal{G}}(\tilde{z})\,,  \\ 
    &  h(z) = k(z) = e^{2\tilde{\eta}(\tilde{z})}\,, \\  
    & f(z) = 1\,.
\end{split}      
\end{equation}
From Eq.~(\ref{eqn:vbfinal}), the butterfly velocity from the entanglement wedge method is obtained as
\begin{eqnarray}
v_{B}^{2} & = & \frac{\tilde{\mathcal{G}}'(\tilde{z}_h)}{6 \tilde{\eta}'(\tilde{z}_h) }  = \frac{1}{6 \eta'(z_h) \mathcal{G}_{0}^{\text{far}}}\,.
\label{vbewmodel4}
\end{eqnarray}
Here we have written down the butterfly velocity expression in terms of numerically generated black hole backgrounds by converting standard coordinates into numerical coordinates.

\begin{figure}[h]
    \centering
\includegraphics[width=0.75\textwidth]{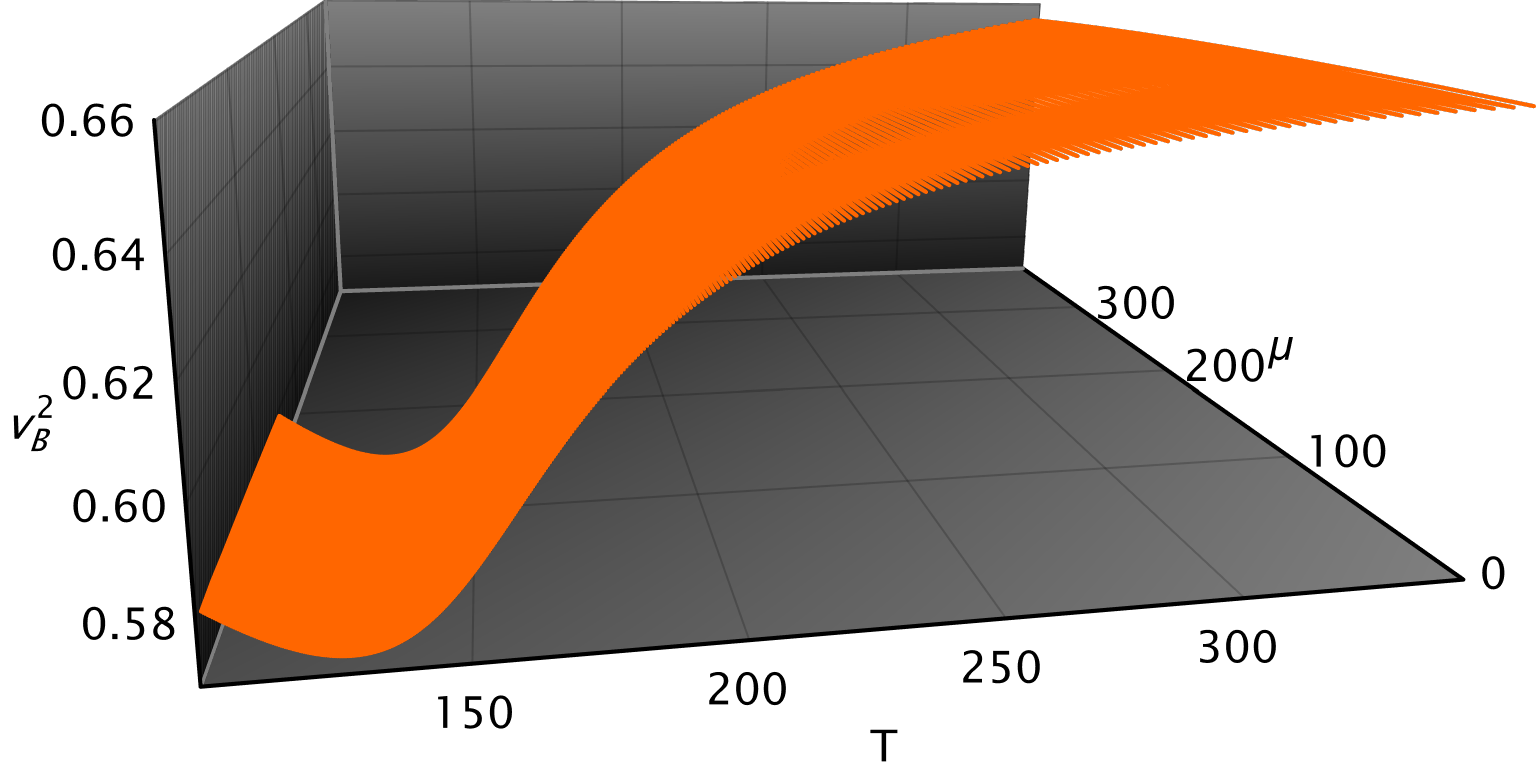}
    \caption{Variation of $v_B$ with respect to $\mu$ and $T$. In units of MeV. }
    \label{fig:TvsmuvsVbmodel4}
\end{figure}

The numerical results of the butterfly velocity as a function of temperature and chemical potential are shown in Figure~\ref{fig:TvsmuvsVbmodel4}. To obtain this result, we numerically constructed approximately  $\sim 5\times 10^5$ charged black hole solutions across a two-dimensional grid of initial conditions $\{\phi_0, \Phi_1\}$. The parameter $\phi_0$ was varied in uniform increments from $0.55$ to $5.5$, while for each value of $\phi_0$, the ratio $\Phi_1/\Phi_{1}^{\text{max}}(\phi_0)$ was varied from $0$ to $0.5$ in equally spaced steps.\footnote{For this system, the maximum value $\Phi_1$ can take is bounded by $\phi_0$ and is given by $\Phi_1 < \Phi_{1}^{max}  \sqrt{-\frac{2 V(\phi_0)}{f(\phi_0)}}$.} We again remind the reader that $\phi_0$ corresponds to the value of the scalar field at the horizon and $\Phi_1$ is the electric field at the horizon. We find that for relatively large temperatures, $T>200~\text{MeV}$, $v_B$ monotonically increases with the temperature, and then, as in the case of the other three models, it asymptotes to the conformal chargeless plasma value in the infinite-temperature limit. However, as opposed to the other three models, $v_B$ now also shows a nonmonotonic variation at relatively low temperature $T<130~\text{MeV}$, i.e., $v_B$ decreases with temperature for $T<130~\text{MeV}$.

At this point, it is important to keep in mind that, unlike the model considered in Section~\ref{subsec:ourmodel}, there is no Hawking/Page phase transition in the gravity side in the current holographic model. This hinders the determination of the dual deconfinement transition temperature via direct methods in this model. To get the deconfinement temperature, one therefore relies on indirect methods, such as the temperature at which the speed of sound attains a minimum value. From this indirect method, the deconfinement transition temperature can be approximated to be around $T_c\sim 140~\text{MeV}$ \cite{Rougemont:2015wca}. Therefore, it is safe to say that the butterfly velocity again exhibits monotonic thermal structure in the high-temperature deconfined phase.

Furthermore, interestingly, just like in the case of the other three models, $v_B$ monotonically decreases with the chemical potential for all fixed values of the temperature. This observation further suggests a universal behavior in the rate of information propagation, associated with the spread of perturbations and operator growth, within the local degrees of freedom of the dual deconfined plasma, which decreases with increasing chemical potential.  Note that, due to the marginal decrease in $v_B$ with respect to $\mu$, this fact is not clearly evident from Figure~\ref{fig:TvsmuvsVbmodel4}, but one can verify this by obtaining a two-dimensional plot of $v_B$ vs $\mu$ at various fixed values of $T$. 

We conclude this section by highlighting an intriguing similarity in the thermal profiles of the butterfly velocity and the speed of sound $c_s$ within this holographic model. Both quantities exhibit a qualitatively similar nonmonotonic dependence on temperature: they decrease at low temperatures, attain a minimum around an intermediate scale, and subsequently increase, saturating to a chemical potential independent value in the high-temperature regime. This asymptotic behavior is naturally understood in terms of the emergent scale invariance that characterizes the ultraviolet (UV) fixed point of the dual field theory. In this limit, the geometry approaches AdS asymptotically, and both transport and chaos-related observables reflect the conformal nature of the underlying theory.

However, the behavior at low temperatures, particularly near the deconfinement temperature,  is less transparent. Unlike in the UV, there is no general symmetry argument that dictates the observed similarity between the butterfly velocity and the speed of sound in the infrared (IR). We also analyze other numerical holographic QCD models of the type \cite{Rougemont:2015wca}, such as \cite{DeWolfe:2010he} and \cite{Critelli:2017oub}, and find analogous thermal behavior between $v_B$ and $c_s$. This apparent parallel thermal behavior may suggest deeper dynamical connections or shared sensitivity to the underlying nonperturbative physics of the strongly coupled plasma, though further investigation is required to elucidate the mechanisms responsible for this parallel behavior.

\subsection{$v_{B}$ from the OTOC method}
Substituting the metric functions \eqref{metricmodel4tilde} and \eqref{eqn:zhminusuv} into \eqref{eqn:relationofAandB}, we get\footnote{Here, to avoid any confusion, we use the tilde on Kruskal coordinates $u$ and $v$ as well.}
\begin{eqnarray}
A(\tilde{u},\tilde{v}) & = & \frac{\tilde{\mathcal{G}}(\tilde{z}_h - \tilde{u} \tilde{v}) e^{2 \tilde{\eta}(\tilde{z}_h - \tilde{u} \tilde{v})}}{2 \alpha^2 \tilde{u} \tilde{v}}\,, \nonumber \\
B(\tilde{u},\tilde{v}) & = & e^{2 \tilde{\eta}(\tilde{z}_h - \tilde{u} \tilde{v})}\,.
\end{eqnarray}
Computing the near-horizon limits of $A$ and $\partial_{\tilde{u}}\partial_{\tilde{v}}B$ as $\tilde{u}\rightarrow 0$, we arrive at the following:
\begin{eqnarray}
    A(0) & = & \lim_{\tilde{u} \to 0}  A(\tilde{u},\tilde{v}) =  -\frac{\tilde{\mathcal{G}}'(\tilde{z}_h) e^{2 \tilde{\eta}(\tilde{z}_h)}}{2 \alpha^2}\,, \nonumber \\
\partial_{\tilde{u}}\partial_{\tilde{v}}B(0) & = &\lim_{\tilde{u} \to 0} \partial_{\tilde{u}}\partial_{\tilde{v}}B(\tilde{u},\tilde{v}) = -2 \tilde{\eta}'(\tilde{z}_h) e^{2\tilde{\eta}(\tilde{z}_h)}\,.
\end{eqnarray}
Substituting the above equations into (\ref{eqn:VBOTOC}) and simplifying, we get the butterfly velocity from the OTOC method as
\begin{eqnarray}
v_{B}^{2} & = & \frac{\tilde{\mathcal{G}}'(\tilde{z}_h)}{6 \tilde{\eta}'(\tilde{z}_h) }  = \frac{1}{6 \eta'(z_h) \mathcal{G}_{0}^{\text{far}}}\,,
\label{vbOTOCmodel4}
\end{eqnarray}
where we have used the fact that $\tilde{\eta}'(\tilde{z}_h)=\eta(z_h)=0$. Comparing (\ref{vbOTOCmodel4}) with (\ref{vbewmodel4}), we again see that the expression of $v_{B}^{2}$ takes the same form in both entanglement wedge and OTOC methods.

\subsection{$v_{B}$ from the pole-skipping method}
Although analytic results for the energy-momentum tensor are difficult to obtain in the holographic QCD model considered in this section, it is straightforward to see that, as this model is diffeomorphism invariant and falls in the EMD gravity definition of type (\ref{eqn:EMDaction}), the energy-momentum tensor again satisfies the condition \eqref{eqn:RHSgeneralperturbEE}. Accordingly, the pole-skipping points are again given by Eq.~(\ref{eqn:TtermPSpointcomputated}). Explicitly, after a little bit of algebra, they take the following form:
\begin{eqnarray}
\omega_{*} & = & 2\pi i T = \frac{i}{2\phi_A^{1/\nu} \sqrt{\mathcal{G}_0^{\text{far}}}} \Lambda \,, \\
q_* & = & i \sqrt{6 \pi T \tilde{\eta}'({\tilde{z}_h}) e^{\tilde{\eta}({\tilde{z}_h})}}   = i \sqrt{\frac{3 \eta'(z_h)}{2 \phi_A^{2/\nu}}} \Lambda\,.
\end{eqnarray}
Again, the butterfly velocity from the pole-skipping points
\begin{eqnarray}
v_{B}^{2} & = &  \frac{\omega_{*}^2}{q_{*}^2} =  \frac{1}{6 \eta'(z_h) \mathcal{G}_{0}^{\text{far}}}\,
\end{eqnarray}
matches exactly with the expression obtained using the entanglement wedge and OTOC methods.

\begin{figure}[h]
    \centering
    \begin{subfigure}[b]{0.45\textwidth}
        \centering
\includegraphics[width=\textwidth]{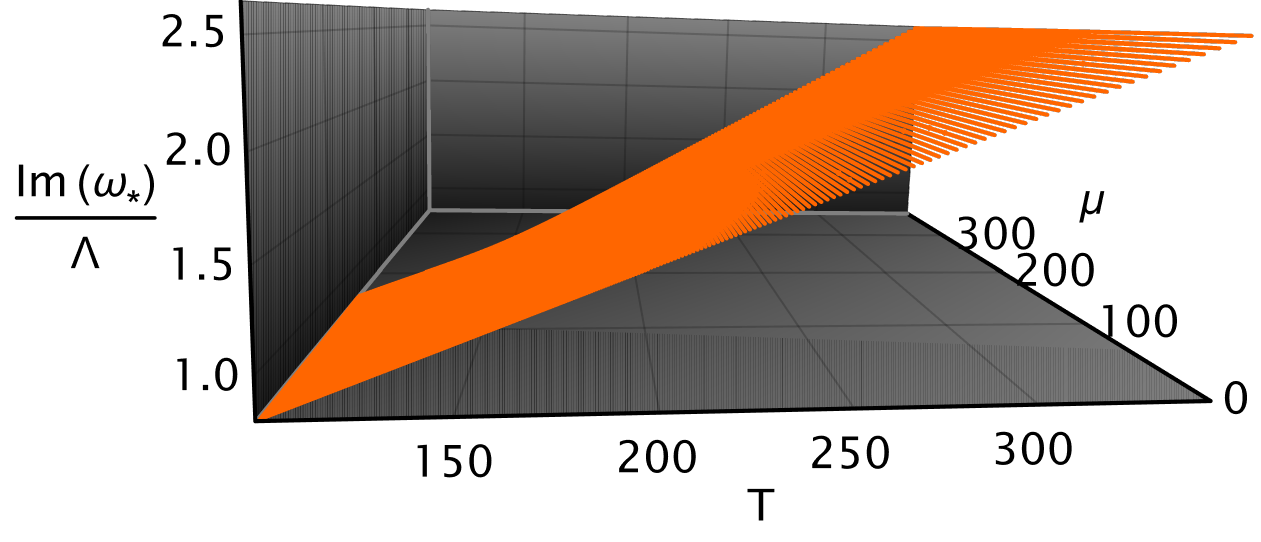}
\label{fig:Tvsmuvsomegastarmodel4}
    \end{subfigure}
    \begin{subfigure}[b]{0.45\textwidth}
        \centering
\includegraphics[width=\textwidth]{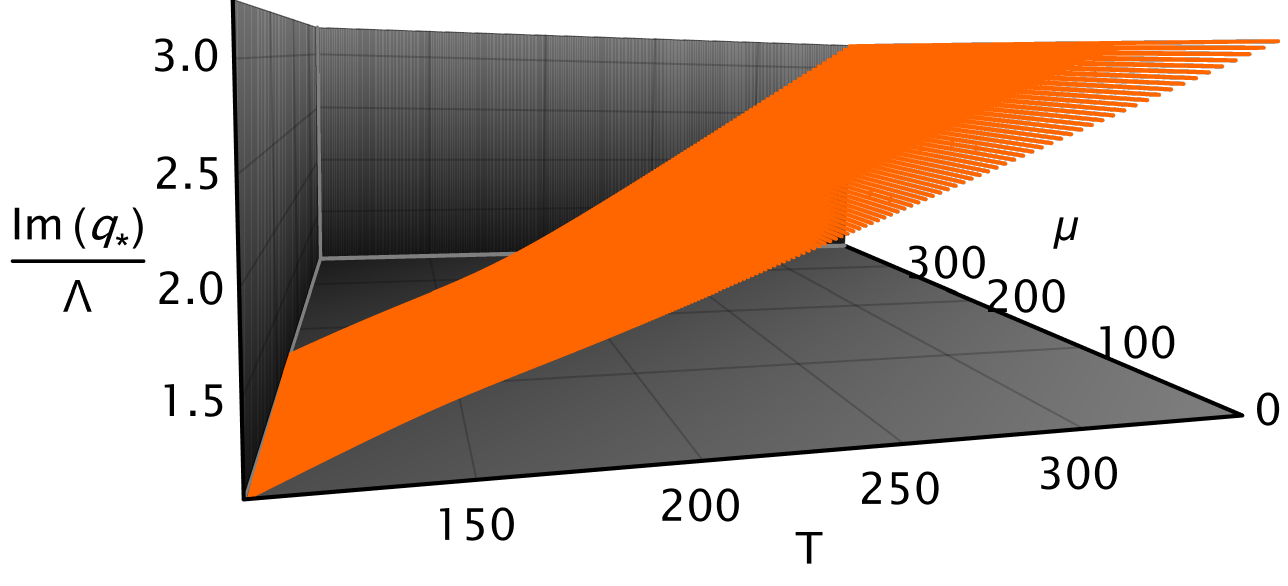} 
    \label{fig:Tvsmuvsqstarmodel4}
    \end{subfigure}
    \caption{Pole-skipping points $\omega_*$ and $q_*$ as a function of temperature and chemical potential. In units of MeV. }
    \label{fig:combinedmodel4}
\end{figure}

It is also interesting to analyze the structure of the pole-skipping points in this model. The results are shown in Figure~\ref{fig:combinedmodel4}, where the imaginary parts of $\omega_{*}$ and $q_{*}$ as a function of $\mu$ and $T$ are plotted. The overall behavior is similar to that of Figure~\ref{fig:combined}. Again, these special points are smooth in the parameter space of temperature and chemical potential, with minimal dependence on the latter.

We conclude by noting a technical subtlety in the holographic model considered in \cite{Rougemont:2015wca}. The last term in the gauge-kinetic function $\tilde{f}(\phi)$, as defined in Eq.~(\ref{potentialmodel4}), 
becomes effectively negligible for most values of $\phi$ due to the large numerical value of the parameter $b_3$, contributing meaningfully only in the limit of very small $\phi$. This exponential term is included to ensure that $\tilde{f}(0)=1$. However, this choice also influences the scaling of the chemical potential and charge, particularly when fitting to susceptibility data [from which the parameters of $\tilde{f}(\phi)$ are fixed], and should therefore be handled with care.  Enforcing $\tilde{f}(0)=1$ results in a spurious narrow spike in the function $\tilde{f}(\phi) $ at small $\phi$. This issue has been discussed in great detail in \cite{Jokela:2024xgz}, where it was further pointed out that the physical impact of this spike is essentially negligible.

An alternative approach, explored in~\cite{Jokela:2024xgz}, involves constructing a holographic QCD model without imposing the condition $\tilde{f}(0)=1$. In this model, $\tilde{f}(\phi)$, like the potential function $V(\phi)$, was taken to have a polynomial profile. We carried out a complementary analysis of the butterfly velocity in this setup and found that it displays qualitative features consistent with those observed earlier, thereby reinforcing the universality of the butterfly velocity across different holographic QCD models.

\section{Conclusion and Future Outlook}
\label{sec:conclusion}
In this work, we conducted a detailed investigation of the butterfly velocity in the deconfined plasma phase of EMD-inspired holographic QCD models. Our analysis focused on four distinct yet physically motivated models that span both top-down and bottom-up approaches within the holographic framework. Specifically, we examined two top-down constructions -- the 1RCBH and 2RCBH models -- derived from consistent truncations of type IIB supergravity. In addition, we analyzed a semianalytic bottom-up model based on the potential reconstruction technique, which allows for analytic control over the metric functions and facilitates matching to QCD-like thermodynamics. Finally, we considered a fully numerical bottom-up model, designed to reproduce lattice QCD thermodynamic data at finite temperature and chemical potential, thereby providing a phenomenologically grounded dual description. This diverse set of models enables us to probe the robustness of chaotic transport properties in strongly coupled nonconformal plasmas, particularly in relation to the butterfly velocity and its thermal- and chemical-potential-dependent behavior in QCD.

In each of these holographic models, we computed the butterfly velocity using three distinct methods: entanglement wedge reconstruction, OTOCs, and pole skipping, and found that all three give the same result. This extends previous results by demonstrating that such an equivalence holds in a variety of holographic QCD models as well. In the first three models, we obtained results analytically, while in the fourth model, we obtained results numerically. We further analyzed the behavior of $v_B$ with respect to the chemical potential and temperature and observed a universal trend across models. In particular, $v_B$ is found to be decreasing/increasing with the chemical potential/temperature. The same result is also true in the simplest holographic charged plasma, i.e., dual to the Reissner-Nordstr\"{o}m AdS background (see the Appendix~\ref{subsec:RNADS}). Such universal behavior offers potential signatures of quantum chaos in QCD that could be accessible to experimental observation, especially in light of the recently proposed experiments to measure OTOCs in quantum systems. Such universal chaotic features might also provide interesting insights into QCD. Note that $v_B$ was also studied recently for non-Abelian gauge theories at high temperatures ($2 - 10~T_c$) using lattice techniques, and it was found that the butterfly velocity does not vary significantly compared to its value at the deconfinement transition temperature \cite{Guin:2025lpy}. In particular, a variation of about $10\%$ at higher temperatures relative to the deconfinement temperature was observed. This can be compared to the results of Section~\ref{subsec:ourmodel}, where the deconfinement temperature can be consistently defined and a variation of $25\%$ is found. Similarly, a variation of $8\%$ is found as we go from $T=150$ to $T=500~\text{MeV}$ in the model of Section~\ref{subsec:Bottomu2p}.

Another important remark that we would like to make is regarding the behavior of $v_B$ with respect to the charge. In this work, we primarily focused on studying the variation of $v_B$ with respect to the chemical potential, but one can also study the variation with respect to the charge. On analyzing, one finds that $v_B$ monotonically decreases even with the charge for all the models presented in our work. It is interesting to note that such a monotonically decreasing trend with charge was also observed in (hyperscaling-violating) Lifshitz theories \cite{Lilani:2024fth}. Thus, it would be interesting to investigate if this trend is a universal feature of all holographic theories or all quantum systems, in general. 

In this work, we restricted our attention to bulk black hole geometries that are isotropic and planar. One can also extend this analysis to black hole geometries that are anisotropic. The work of \cite{Huang:2018snb} provides an example to compute butterfly velocity in anisotropic backgrounds. Exploring the effect of magnetic fields on the butterfly velocity in holographic QCD models is another interesting avenue. However, since in most holographic models magnetic field induces an anisotropy, computing the butterfly velocity may not be an easy task. One exception to this is the model discussed in \cite{Wang:2024bli}. This model has a $SO(3)$ symmetry in spite of the magnetic field. For this model, one can check, using the methods discussed in this work, that the butterfly velocity monotonically decreases as a function of the magnetic field. It would be interesting to investigate if this decreasing trend is also a feature of other more sophisticated holographic models of magnetized QCD \cite{Bohra:2020qom,Jena:2024cqs}.

Before concluding, we would like to make a brief remark on the interplay between the butterfly velocity and phase transitions. In~\cite{Baggioli:2018afg}, it was conjectured that $v_B$ can serve as a probe of phase transitions. That work presented examples within AdS/CMT models supporting this idea. The bottom-up holographic model considered in Section~\ref{subsec:ourmodel} exhibits a Hawking/Page phase transition. Notably, we have plotted $v_B$ only for the thermodynamically stable large black hole branch. However, as mentioned earlier, there also exists a thermodynamically unstable small black hole branch, which appears only below a critical chemical potential and vanishes above it. Extending the plot of $v_B$ to include this unstable small branch reveals a nonmonotonic behavior for values of the chemical potential below the critical point. In contrast, for values above the critical chemical potential, $v_B$ exhibits monotonic behavior once again. This suggests that the onset of nonmonotonicity coincides with the critical chemical potential, providing further evidence in support of the conjecture put forward in~\cite{Baggioli:2018afg}. Note that, by taking different profiles of the scale factor $A(z)$, one can also construct gravity solutions exhibiting van der Waals type first-order small/large black hole phase transitions with a second-order critical point \cite{Dudal:2017max}. It would be interesting to see if information about such transitions and critical points can also be reflected in the butterfly spectrum. There are already interesting suggestions that such small/large phase transitions can be probed by the Lyapunov exponents \cite{Shukla:2024tkw, Guo:2022kio}. Work in this direction is under progress.

\section*{Acknowledgements}
The work of S.M.~is supported by the core research grant from the Science and Engineering Research Board, a statutory body under the Department of Science and Technology, Government of India, under grant agreement number CRG/2023/007670.

\section*{Data Availability}
The data are available from the authors upon request.

\appendix

\section{CFT dual to planar RN-AdS black hole.}
\label{subsec:RNADS}
\subsection{Background}
In this section, we compute $v_{B}$ of the charged plasma corresponding to the Reissner-Nordstr\"{o}m AdS (RN-AdS) black hole background. The metric of an RN-AdS black hole in five dimensions is
\begin{equation}
    \label{eqn:SYMmetric1}
    ds^{2} = z^{2}\left[-F(z)dt^{2}+\frac{1}{F(z) z^{4}}dz^{2} + dx^{i}dx^{i}\right]\,.
\end{equation}
Here $F(z)$ is the blackening function given by (in units of $L=1$)
\begin{equation}
    \label{eqn:FzSYM}
    F(z) = 1 - (1+Q^{2})\left(\frac{z_h}{z}\right)^{4} + Q^{2}\left(\frac{z_h}{z}\right)^{6}\,.
\end{equation}
The chemical potential and temperature are
\begin{equation}
    \label{eqn:muandTSYM}
    \mu = \sqrt{3} Q z_h\,,~~~ T = \frac{z_h}{\pi}\left(1-Q^{2}/2\right)\,.
\end{equation}

\subsection{Calculation of $v_{B}$ using entanglement wedge method}
Comparison of \eqref{eqn:SYMmetric1} with \eqref{eqn:generalmetric} yields the following:
\begin{equation} 
\label{eqn:metricfunctionsRNAdS}
  g(z) = F(z)\,,~~~ h(z) = k(z) = z^{2}\,,~~~ f(z) = \frac{1}{z^2}\,.
\end{equation}
Substituting the horizon values of the metric functions (and their derivatives) into \eqref{eqn:vbfinal}, we get the following expression for the butterfly velocity:
\begin{equation}
    \label{eqn:vbSYM}
    v_{B}^{2} = \frac{2}{3}\left(1-\frac{Q^{2}}{2}\right)\,.
\end{equation}
Solving for $Q$ from \eqref{eqn:muandTSYM}, we get (considering only the positive root)
\begin{equation}
  \label{eqn:chargeRNAdS}  
  Q (\mu, T) = \frac{\sqrt{2 \mu ^2+3 \pi ^2 T^2}-\sqrt{3} \pi  T}{\mu }\,.
\end{equation}
Substituting \eqref{eqn:chargeRNAdS} into \eqref{eqn:vbSYM}, we get the following expression for the butterfly velocity in terms of $\mu$ and $T$:
\begin{equation}
    \label{eqn:vBRNAdS}
    v_B^2 = \frac{2 \pi  T \left(\sqrt{6 \mu ^2+9 \pi ^2 T^2}-3 \pi  T\right)}{3 \mu ^2}\,.
\end{equation}

\begin{figure}[t]
    \centering
    \begin{subfigure}[b]{0.4\columnwidth}
        \centering
\includegraphics[width=1.15\textwidth]{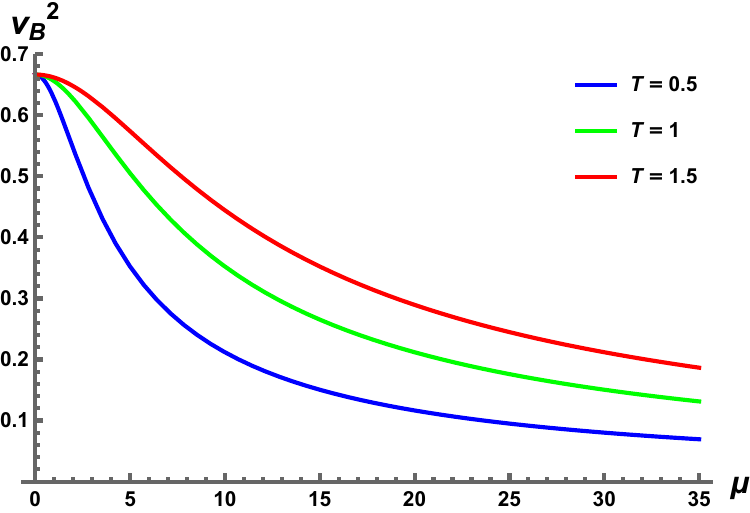} 
        \caption*{(a) $v_B^2$ vs $\mu$ at various fixed values of $T$.}
    \end{subfigure}\hspace{10mm}
    \begin{subfigure}[b]{0.4\columnwidth}
        \centering
\includegraphics[width=1.15\textwidth]{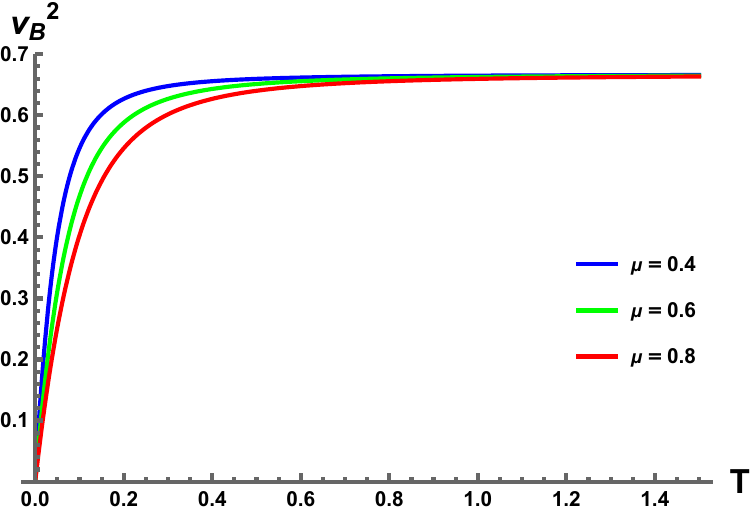} 
        \caption*{(t) $v_B^2$ vs $T$ at various fixed values of the chemical potential.}
    \end{subfigure}\hspace{10mm}
    \caption{Variation of $v_B$ with respect to $\mu$ and $T$ for RN-AdS background.}
    \label{fig:RNADSmodel}
\end{figure}

Figure~\ref{fig:RNADSmodel} illustrates the behavior of $v_B$ with the RN-AdS background. As shown, the qualitative behavior of $v_B$ in this setup closely mirrors the trends observed in the other holographic models discussed in previous sections.

\subsection{Calculation of $v_{B}$ using the OTOC method}
Comparing \eqref{eqn:relationofAandB} and \eqref{eqn:FzSYM}, we get
\begin{equation}
    \label{eqn:AandBforSYM}
    A(u,v) = \frac{\left[1 - (1+Q^{2})\left(\frac{z_h}{z_h-uv}\right)^4 + Q^{2}\left(\frac{z_h}{z_h-uv}\right)^6\right]\left(z_h-uv\right)^2}{2 \alpha^{2}uv}\,,~~~ B(u,v) = (z_h - uv)^2\,.
\end{equation}
The near-horizon limits are
\begin{equation}
    \label{eqn:AHorizonSYM}
    A(0)= \frac{(Q^{2}-2)z_h}{\alpha^{2}}\,,~~~ \partial_{u}\partial_{v}B(0) = - 2 z_h\,.
\end{equation}
Putting these near-horizon values into \eqref{eqn:VBOTOC}, we get the same result as \eqref{eqn:vbSYM}.

\subsection{Calculation of $v_{B}$ using pole-skipping method}
Substituting the values of metric functions in \eqref{eqn:TtermPSpointcomputated}, we get the pole-skipping points
\begin{equation}
    \label{eqn:PSpointSYM}
    \omega_{*}^2=-z_{h}^2(-2 + Q^2)^2\,,~~~ \hspace{0.3cm} q_{*}^2=-z_{h}^2( 6 - 3 Q^2)\,.
\end{equation}
The expression for the butterfly velocity  $v_{B}^{2} =\omega_{*}^2/q_{*}^2$, 
\begin{equation}
    \label{eqn:vbSYMPS}
    v_{B}^{2} =\frac{\omega^2}{q^2}=\frac{2}{3}\left(1 - \frac{Q^2}{2}\right)\,,
\end{equation}
again matches with the other two methods.


\bibliography{biblio.bib}
\bibliographystyle{JHEP}

\end{document}